\documentclass[aps,prb,twocolumn]{revtex4}
\usepackage{xcolor,soul}

\usepackage{graphicx}
\usepackage{placeins}
\usepackage{blindtext}
\usepackage{enumitem}
\usepackage{amsfonts}

\usepackage{amsmath,latexsym,amsthm}

\begin{document}

\setlength{\fboxsep}{0pt}
\setlength{\fboxrule}{.1pt}

\title{Equality of ADM mass and generalized Komar energy\\
in asymptotically-flat dynamical spacetimes}

\author{Zhi-Wei Wang}
\email{zhiweiwang.phy@gmail.com}
\affiliation{College of Physics, Jilin University, Changchun,
130012, People's Republic of China}

\author{Samuel L. Braunstein}
\email{sam.braunstein@york.ac.uk}
\affiliation{University of York, York YO10 5GH,
United Kingdom}


\begin{abstract}

\noindent

We find a relation between the ADM mass and a generalized Komar energy
in asymptotically-flat spacetime. We do not need to assume the existence
of either a Killing or even asymptotically-Killing vector field.
Instead, our generalized Komar energy is constructed from the normal
evolution vector (the lapse function times the future-directed unit
normal to the spacelike hypersurfaces on which the ADM mass is
measured). We find equality between the ADM mass and this generalized
Komar energy even for dynamical asymptotically-flat spacetimes provided
the 3-dimensional Einstein tensor drops off quickly enough at spatial
infinity, in particular, whenever ${}^{(3)}G_{ij}=o(r^{-3})$. No
additional assumptions are required for equality. As this generalized
energy is fully covariant, it may provide a powerful tool for analyzing
energy content in dynamical spacetimes containing compact objects.


\end{abstract}

\maketitle


\section{Introduction}

The definition and interpretation of energy within the framework of
general relativity is one of its most profound and enduring challenges.
Unlike in flat spacetime or theories where gravity is treated as a field
within a fixed background, the absence of a global inertial frame and
the equivalence principle mean that gravitational field energy cannot be
localized to specific points in spacetime in a diffeomorphism-invariant
way.\cite{Wald1984} However, significant progress has been made in
defining energy as a global quantity at spatial or null infinity with
reasonable geometric constraints.

In particular, in 1959, Arnowitt, Deser, and Misner (ADM), utilizing the
3+1 decomposition of spacetime, developed the Hamiltonian formalism.
\cite{Arnowitt1959, Arnowitt1960, Arnowitt1962} Their analysis for
asymptotically-flat spacetimes led to the definition of the ADM mass.
This represents the total energy of an isolated dynamical system over a
spacelike hypersurface. The ADM mass is a global quantity, well-defined
for asymptotically-flat dynamical initial data surfaces. Another crucial
concept for dynamical spacetimes is the Bondi mass, \cite{Bondi62}
defined at future null infinity, however this quantity is beyond the
scope of the study of this paper.

On the other hand, energy can also be defined as a conserved quantity
conjugate to time translation symmetry via Noether's theorem. A
well-known example is the Komar mass,\cite{Komar1959} defined for
stationary spacetimes via a covariant surface integral involving the
timelike Killing vector field, which is the generator of time
translation symmetry. The Komar integral provides a conserved quantity
related to the Noether ``charge'' associated with this spacetime symmetry.
For asymptotically-flat stationary spacetimes, the Komar mass calculated
using the asymptotically-timelike Killing vector is known to coincide
with the ADM mass.\cite{beig1978} However, it is also thought they are
only equal in a stationary spacetime since the traditional Komar mass
definition is only for stationary spacetimes.

A central challenge in extending the symmetry-dependent Komar mass to
general dynamical spacetimes is the absence of global Killing vector
fields. This has motivated research into ``generalized'' Komar
expressions by replacing the Killing vector by another vector field
chosen based on some physical or geometric
criteria.\cite{komar1962,komar1963,harte2008,feng2018,wang2020,wang2021}
Komar himself made a seminal contribution in this area by introducing
asymptotic Killing vector fields, e.g., the semi-Killing vector and the
almost-Killing vector.\cite{komar1962,komar1963} He first argued that
the semi-Killing vector field should allow one to define a generalized
Komar energy on an asymptotically-flat hypersurface, even one containing
gravitational waves.\cite{komar1962} He then argued that to ensure this
generalized Komar energy is a reasonable generalization of the energy in
asymptotically Lorentz-covariant theories, this vector field needs to be
almost Killing.\cite{komar1963} Since these asymptotic Killing vectors
must be orthogonal to the spacelike hypersurface, the generalization of
these Killing vectors corresponds to a selection of the asymptotic
conditions of the hypersurfaces.\cite{komar1963}

To transcend the asymptotic flatness-constraints, Harte replaced the
Killing vector with generalized affine collineations constructed locally
around a specific observer's worldline.\cite{harte2008} This
observer-dependent vector makes the generalized Komar energy and momentum
not an intrinsic property of the spacetime region but quasilocal and
non-conserved quantities. Harte interpreted the rate of mass change as
matter flux or `gravitational current'.\cite{harte2008} To overcome the
dilemma of Komar current non-conservation caused by radiation energy,
Feng constructed some new global conserved (Komar) currents based on
various generalized `Killing' vectors and scalar test
fields.\cite{feng2018} By analyzing the outgoing Vaidya spacetime, Feng
demonstrated that such generalized Komar currents can yield conserved
quantities behaving as expected for radiated energy.\cite{feng2018}



Although the above studies have pushed the application of Komar energy to
much more generic scenarios, like asymptotically-flat dynamical
spacetimes, proving a general equivalence between the global ADM mass
and a generalized Komar integral in dynamical spacetimes has remained a
significant challenge. Specifically, since the dynamical spacetime
metric is time-dependent and asymmetric, we need both to suitably
generalize the original Komar mass in the absence of any Killing vector
and to prove its equality with the ADM mass. Here, we construct a particular
asymptotically-timelike vector field that plays the role of the Killing
vector in a Komar-like integral. This vector field is not a global
Killing vector field, but it is defined based on the asymptotic
structure of the spacetime. We then rigorously prove that the ADM mass
precisely equals to the Komar-like form integrated over a surface at
spatial infinity under reasonable asymptotically-flat conditions. This
finding extends the known ADM=Komar equality, previously established for
stationary, symmetric spacetimes, to a broad class of asymptotically-flat
dynamical spacetimes. 

The remainder of this paper is organized as follows. In Section~II, we
define the generalized Komar energy for an arbitrary vector field
$\xi^\mu$ and briefly review the 3+1 split formalism. In Section~III, we
review the standard asymptotically-flat conditions and in Section~IV,
the ADM mass based on this. Section~V is the core of our analysis, where
we present the detailed proof of equality between the ADM mass and our
generalized Komar energy $E(\xi)$ for asymptotically-flat dynamical
spacetimes. This involves selecting a specific asymptotically-timelike
vector field $\xi^\mu$ that approaches a time translation at infinity
and meticulously transforming the ADM mass integral, demonstrating its
equivalence to the generalized Komar integral under the derived
asymptotic conditions. In Section~VI, we explicitly consider the
conservation of the generalized Komar energy and provide a summary of
our findings. Throughout this work we set $G=c=\hbar=k_B=1$, and we
suppose that the spacetime is asymptotically-flat from Section III
onwards. Greek indices run from 0 to 3, and lower-case Latin indices run
from 1 to 3 and when used, upper-case Latin indices run from 2 to 3.

\section{Construction of the Conserved Komar Energy-Momentum }

Here we review Komar's approach to define a conserved
energy-momentum even on dynamical spacetimes.
\cite{Komar1959,komar1962,komar1963} For an arbitrary
vector field $\xi^\mu$, we introduce the antisymmetric tensor
$S^{\mu\nu}(\xi)$, where
\begin{eqnarray}
 S^{\mu\nu}(\xi ) \equiv \frac{1}{2} (\xi^{\nu;\mu}
- \xi^{\mu;\nu}) \equiv {\xi^{[\nu;\mu]}} .
 \label{flux0_re}
\end{eqnarray}
Like the anti-symmetric electromagnetic field tensor,
this tensor has a corresponding `energy' density flux vector
$J^\mu(\xi)$ given by \cite{Komar1959}
\begin{eqnarray}
 J^\mu(\xi ) \equiv S^{\mu\nu}(\xi )_{;\nu} = {\xi^{[\nu;\mu]}}_{;\nu} \;.
 \label{flux1_re}
\end{eqnarray}
A key property of $J^\mu(\xi)$ is that its covariant divergence vanishes
identically for any vector field $\xi^\mu$. This can be shown as follows:
\begin{eqnarray}
  {J^\mu}_{;\mu} &=& {{\xi^{[\nu;\mu]}}}_{;\nu\mu} 
  = \frac{1}{2} (\xi^{\nu;\mu}{}_{;\nu\mu} -
\xi^{\mu;\nu}{}_{;\nu\mu}) \nonumber \\ 
  &=& \frac{1}{2} (\xi^{\nu;\mu}{}_{;\nu\mu} -
\xi^{\nu;\mu}{}_{;\mu\nu}) \nonumber \\ 
  &=& \frac{1}{2} (R^\nu{}_{\alpha\nu\mu}\xi^{\alpha;\mu} +
R^\mu{}_{\alpha\nu\mu}\xi^{\nu;\alpha}) \nonumber \\
  &=& \frac{1}{2} R_{\alpha\beta} (\xi^{\beta;\alpha} -
\xi^{\alpha;\beta}) = 0,
  \label{con0_re}
\end{eqnarray}
where the Ricci identity for the commutator of covariant derivatives
acting on a tensor is used in moving from the second to the third line.

\begin{figure}[ht]
\centering
\includegraphics[width=0.25\textwidth]{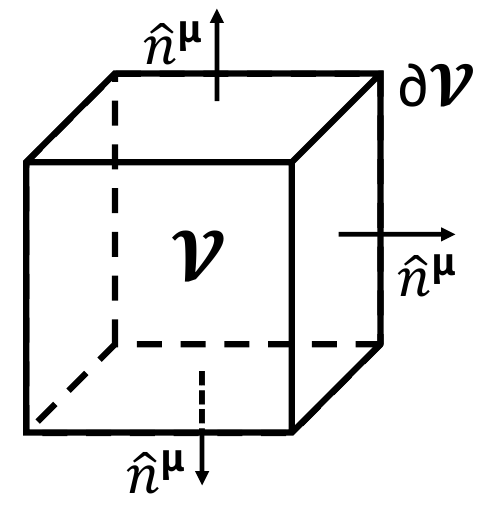} 
\caption{This 4-volume $ {\cal V} $ is a subset of the entire
spacetime manifold ${\cal M}$.  Here $ \partial {\cal V} $ is the
boundary of $ {\cal V}$, and $\hat n^\mu$ is the outgoing unit
vector normal to the boundary $ \partial {\cal V}$. } 
\label{fig1}
\end{figure}

The vanishing divergence, ${J^\mu}_{;\mu} = 0$, implies $J^\mu$ is a
locally covariantly conserved quantity for arbitrary vector fields
$\xi^\mu$. This was Komar's original observation about this
quantity.\cite{Komar1959} Integrating Eq.~(\ref{con0_re}) over an
arbitrary 4-volume $\mathcal{V}$ within the spacetime manifold
$\mathcal{M}$ and applying Stokes' theorem, we obtain
\begin{equation}
 \int_{\mathcal{V}}   {J^\mu}_{;\mu} \sqrt{-g} \, d^4z
= \int_{\partial {\cal V}} 
 J^\mu \hat n_\mu \sqrt{\gamma^{({\partial {\cal V}})}}\,  d^3x = 0,
 \label{inte4_re}
\end{equation}
where $\partial {\cal V}$ is the 3-dimensional boundary of
${\cal V}$, $\hat n_\mu$ is the outward-pointing unit normal to
$\partial {\cal V}$ (see Fig.~\ref{fig1}), and
$\gamma^{({\partial {\cal V}})}$ is the determinant of the induced
metric on $\partial {\cal V}$. This means that the current flux into
the 4-volume is the same as the current flux out. This is a
{\it local\/} conservation law for an {\it arbitrary\/} vector field
even in an arbitrary dynamical spacetime.\cite{Komar1959}

\begin{figure}[ht]
\centering
\includegraphics[width=0.35\textwidth]{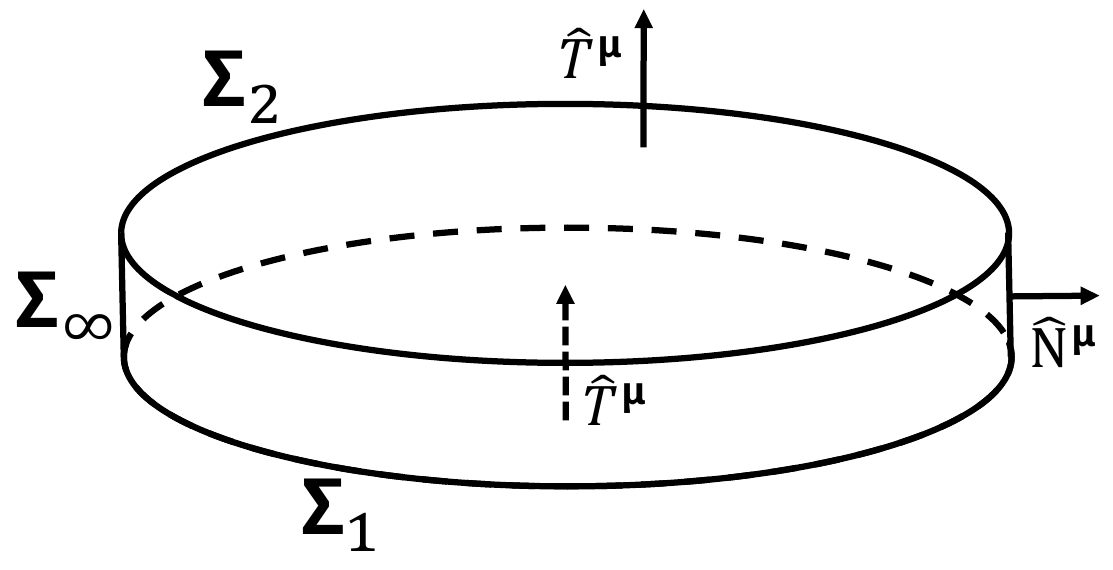}
\caption{This 4-volume $ {\cal V} $ is a region between two infinitely
large three-hypersurfaces $\Sigma_1$, $\Sigma_2$. The boundary
$\partial\mathcal{V}$ is composed of $\Sigma_1$, $\Sigma_2$, and
a timelike boundary at spatial infinity $\Sigma_\infty$. Here
$\hat T^\mu$ is the timelike unit normal vector pointing to the future
on $\Sigma_1, \Sigma_2$, and $\hat L^\mu$ is the spacelike outgoing
unit vector normal to $\Sigma_\infty$. } 
\label{fig2}
\end{figure}

Next, we restrict our attention to dynamical spacetimes that extend to
spatial infinity and are simply connected there. Consider a 4-volume
$\mathcal{V}$ bounded by two spacelike hypersurfaces $\Sigma_1$ and
$\Sigma_2$, and a timelike hypersurface $\Sigma_\infty$ at the unique spatial
infinity (see Fig.~\ref{fig2}). Eq.~(\ref{inte4_re}) implies
\begin{eqnarray}
    \int_{\Sigma_2} J^\mu \hat T_\mu d\Sigma_2 - \int_{\Sigma_1}
J^\mu \hat T_\mu d\Sigma_1 + \int_{\Sigma_\infty}
J^\mu \hat L_\mu d\Sigma_\infty = 0,
 \label{eq:inte41_re}
\end{eqnarray}
where $\hat T_\mu$ is the future-directed timelike unit normal to
$\Sigma_1$ and $\Sigma_2$, and $\hat L_\mu$ is the outward-pointing
spacelike unit normal to $\Sigma_\infty$.
$d\Sigma = \sqrt{\gamma^{(\Sigma)}}\, d^3x$ and $d\Sigma_\infty$ are
the respective volume elements.

If the flux through spatial infinity vanishes, i.e.,
$\int_{\Sigma_\infty} J^\mu \hat L_\mu d\Sigma_\infty = 0$,
then the quantity\cite{Komar1959}
\begin{eqnarray}
   E(\xi ) &\equiv& \frac{1}{4\pi} \int_{\Sigma}
J^\mu(\xi) \hat T_\mu \sqrt{\gamma^{(\Sigma)}}\,  d^3x \nonumber \\
   &=& \frac{1}{4\pi} \int_{\Sigma} {\xi^{[\nu;\mu]}}_{;\nu}
\hat T_\mu \sqrt{\gamma^{(\Sigma)}}\,  d^3x
 \label{inte42_re}
\end{eqnarray}
is conserved,\cite{Komar1959} meaning it is independent of the choice
of spacelike hypersurface $\Sigma$.
Now recalling Stokes' theorem\cite{carroll2004}
for an anti-symmetric tensor $F^{\mu\nu}$ 
\begin{equation}
    \int_\Sigma \hat T_\alpha {F^{\alpha \beta}}_{;\beta}
\sqrt{\gamma^ {(\Sigma)} } d^{n-1}x =\! \int_{\partial \Sigma}\!
F^{\alpha \beta} \hat T_\alpha \hat N_\beta
\sqrt{\gamma^{(\partial \Sigma)} } d^{n-2}y,
    \label{stokes_re}
\end{equation}
where now $\hat N_\beta$ is the outward normal to $\partial\Sigma$, but
is also normal to $\hat T_\alpha$.
Applying this to Eq.~(\ref{inte42_re}) with
$F^{\nu\mu} = \xi^{[\nu;\mu]}$ yields the Komar
energy-momentum\cite{Komar1959} as a surface integral over the boundary at
spatial infinity, $\partial \Sigma$,
\begin{equation}
   E(\xi ) = \frac{1}{4\pi} \int_{\partial \Sigma } \!
\xi^{[\nu;\mu]} \hat N_\nu \hat T_\mu 
    \sqrt{\gamma^{(\partial \Sigma)}   } \,d^2y \; .
 \label{inte43_re}
\end{equation}

Although the Komar integral is well-defined for an arbitrary vector
field $\xi^\mu$ even in dynamical spacetimes, it was originally
formulated by Komar for the case where $\xi^\mu$ is a Killing
vector~\cite{Komar1959}. He subsequently generalized this
energy-momentum definition to spacetimes that admit asymptotic Killing
vectors~\cite{komar1962,komar1963}. 

\def\leavemeout1{
To ensure both the positivity of this
generalized energy and that the spacetime is locally flat when the total
energy vanishes, Komar further showed that for the choice of
$\xi^\mu= (\partial_t)^\mu$, i.e., as the the hypersurface-orthogonal
time-translation vector, the metric
components must satisfy~\cite{komar1963}
\begin{equation}
   g_{ij,0} = g_{i0,0} = 0.
 \label{KomarAsy}
\end{equation}
}

For asymptotically-flat stationary spacetimes, Beig proved in
1978 that
the Komar mass is equivalent to the ADM mass.\cite{beig1978} However,
the relationship between the Komar energy and the ADM mass in 
dynamical spacetimes has remained an open question. Indeed, the prevailing
consensus is that their equivalence holds only for asymptotically-flat
stationary spacetimes. In this work, we demonstrate that the Komar and
ADM masses are in fact equal for a broad class of asymptotically-flat,
dynamical spacetimes. 

\def\leaveout3{
If $\xi^\mu$ is an asymptotic timelike translation, $E(\xi)$ is
interpreted as the generalized Komar energy; if $\xi^\mu$ is an
asymptotic spacelike translation, $E(\xi)$ is the generalized Komar
momentum (up to a
prefactor).\cite{Komar1959,komar1962,komar1963,harte2008,feng2018} This
$E(\xi)$ reduces to the standard Komar mass if $\xi^\mu$ is a timelike
Killing vector $K^\mu$ normalized such that $K^\mu K_\mu \to -1$ at
spatial infinity. We also require that the net $J^{\mu} \hat N_\mu$
vanish at spatial infinity to ensure $J^{\mu} \hat T_\mu$ is preserved
on every hypersurface. 

Since the evaluation of $\hat T_\mu$ and $\hat N_\nu$ often utilizes the
3+1 decomposition of spacetime, we now review this formalism in more
detail \cite{york1979, gourgoulhon2012}. In the 3+1 split formalism, we
foliate spacetime by a family of non-intersecting spacelike
hypersurfaces $\Sigma_t$, labeled by a time function $t$. The
future-directed unit normal to these hypersurfaces is $\hat T^\mu$. The
time evolution vector $(\partial_t)^\mu$ can be decomposed as
\begin{equation}
(\partial_t)^\mu = {\cal N} \hat T^\mu + \beta^\mu \; ,
\label{shif0_re}
\end{equation}
where $\mathcal{N}$ is the lapse function and $\beta^\mu$ is the shift
vector, tangent to $\Sigma_t$ (i.e., $\beta^\mu \hat T_\mu = 0$).
This implies $\hat T_\mu = -{\cal N} \nabla_\mu t$. In coordinates
adapted to the foliation $(t, x^i)$, we have
\begin{equation}
    \hat T^\mu = \Bigl( \frac{1}{{\cal N}},
-\frac{\beta^i}{{\cal N}} \Bigr) \; ,\;\;\;\;
\hat T_\mu = (-{\cal N}, 0, 0, 0) \;.
\label{shif1_re}
\end{equation}
The induced spatial metric on $\Sigma_t$ is
$\gamma_{\mu\nu} = g_{\mu\nu} + \hat T_\mu \hat T_\nu$.
The determinant of the 4-metric is
$\sqrt{-g} = \mathcal{N}\sqrt{\gamma}$, where $\gamma = \det(\gamma_{ij})$.

The conservation of $E(\xi)$ as defined by Eq.~(\ref{inte43_re}) relies
on the vanishing of the flux $J^\mu \hat N_\mu$ at spatial infinity.
We will later show (in Section~VI, that this condition holds for
asymptotically-flat spacetimes satisfying certain common fall-off
conditions. This should also be the requirement of the ADM mass,
since it is conserved across all spacelike hypersurfaces.
}

\def\leavemeout2{
\begin{figure}[ht]
\centering
\includegraphics[width=0.45\textwidth]{pic3.pdf} 
\vspace{0.1in}
\caption{The Penrose diagram illustrates that for asymptotically-flat
spacetimes considered here, gravitational and light radiation propagate
along null infinity, not reaching spatial infinity. This supports the
notion that a suitably defined energy at spatial infinity can be conserved.}
\label{fig3a}
\end{figure}
}

\section{Asymptotically-Flat Spacetimes}

The basic idea of an asymptotically-flat spacetime is that the spacetime
metric takes the form
\begin{equation}
g_{\mu\nu}=\eta_{\mu\nu}+O\Bigl(\frac{1}{r}\Bigr),
\label{asymp0}
\end{equation}
where $\eta_{\mu\nu} \equiv \text{diag}(-1,1,1,1)$, $x^\mu=(t,x^1,x^2,x^3)$,
$r^2 \equiv (x^1)^2+(x^2)^2+(x^3)^2$, 
and $f(r)=O(r^{-m})$ if
there exists {\it some\/} constant $C > 0$ such that
$\vert f(r)\vert \le C \,r^{-m}$
for all sufficiently large $r$.
In other words, asymptotically, the spacetime approaches flat spacetime.

However, in the literature, there are two generic variations between how
temporal and spatial derivatives are considered to behave. We call these
York-lite and Weinberg conditions. Below we take the usual modern convention
that Greek indices run over $\{0,1,2,3\}$,
whereas Latin indices run only over the spatial degrees of freedom, i.e.,
$\{1,2,3\}$.

\subsection{York-lite asymptotic consitions}

York's approach\cite{York1980} to asymptically-flat spacetimes was to
take Eq.~(\ref{asymp0}) and assume further that under spatial
derivatives the metric and extrinsic curvature ((on
hypersurfaces, $\Sigma$, of constant $t$)
behave as
\begin{eqnarray}
g_{\mu\nu,i} &=& O\Bigl(\frac{1}{r^2}\Bigr) ,\;\;
g_{\mu\nu,ij} = O\Bigl(\frac{1}{r^3}\Bigr) ,\;\;\cdots \nonumber \\
K_{ij}&=& O\Bigl(\frac{1}{r^2}\Bigr) ,\;\;\,
K_{ij,k} = O\Bigl(\frac{1}{r^3}\Bigr) ,\;\; \ldots 
    \label{Yorklite} 
\end{eqnarray}
Since 
\begin{eqnarray}
    K_{\mu\nu} = -\frac{1}{2}\mathfrak{L }_{\hat T} \gamma_{\mu\nu}
 = -\frac{1}{2} \gamma_{\mu\nu ,\alpha} \hat T^\alpha
-\frac{1}{2} {\hat T^\alpha}_{,\mu} \gamma_{\alpha \nu}
-\frac{1}{2} {\hat T^\alpha}_{,\nu} \gamma_{\mu \alpha} ,
    \nonumber
\end{eqnarray}
and $\hat T^\mu=(1,0,0,0)+O(1/r)$, the extrinsic curvature
asymptotically-flat conditions for the extrinsic curvature in
Eq.~(\ref{Yorklite}) reduce to 
\begin{equation}
g_{ij,0}= O\Bigl(\frac{1}{r^2}\Bigr) ,\;\;\;\;
g_{ij,0 k}= O\Bigl(\frac{1}{r^3}\Bigr) , \ldots.
    \label{aymp0}
\end{equation}
Assuming we may reorder derivatives this yields
\begin{equation}
g_{ij,k0} = O\Bigl(\frac{1}{r^3}\Bigr).
\label{gijk0}
\end{equation}
We may now easily calculate that 
\begin{eqnarray}
\Gamma_{\mu \nu i} &=& O\Bigl(\frac{1}{r^2}\Bigr), \nonumber \\
\Gamma_{000}&=& O\Bigl(\frac{1}{r}\Bigr),~~~
\Gamma_{i00}=O\Bigl(\frac{1}{r}\Bigr).
\label{ChrisCond}
\end{eqnarray}
Finally, from Eqs.~(\ref{gijk0}) and~(\ref{ChrisCond}) we find for
the Ricci curvature
\begin{eqnarray}
R_{0i} & =& O\Bigl(\frac{1}{r^3}\Bigr), \nonumber \\
R_{00} &=& O\Bigl(\frac{1}{r^2}\Bigr),~~~R_{ij} = O\Bigl(\frac{1}{r^2}\Bigr).
\label{RicciCond}
\end{eqnarray}
Thus, consistency with the Einstein field equations would suggest that
the energy momentum must satisfy
\newline
\begin{eqnarray}
T_{0i} & =& O\Bigl(\frac{1}{r^3}\Bigr), \nonumber \\
T_{00} &=& O\Bigl(\frac{1}{r^2}\Bigr),~~~T_{ij} = O\Bigl(\frac{1}{r^2}\Bigr),
\label{TmunuCond}
\end{eqnarray}
where we used the fact that $g_{0i}=O(r^{-1})$.
Of course, these are generic conditions based solely on the asymptotic
behavior of the metric and extrinsic curvature; it is mathematically
consistent for the energy-momentum (and hence Ricci curvature) to
actually fall to zero more rapidly.

We call these conditions ``York-lite'' because we do not include in them
his stronger assumptions about how the energy-momentum tapers off
asymptotically.

\subsection{Weinberg asymptotic conditions}

Weinberg\cite{Weinberg1972} took a more covariant approach in expressing
the conditions for a spacetime to be asmptotically flat, namely that
\begin{eqnarray}
    &&g_{\mu\nu}= \eta_{\mu\nu} + O\Bigl(\frac{1}{r}\Bigr) ,\;\;
g_{\mu\nu,\alpha} = O\Bigl(\frac{1}{r^2}\Bigr) ,\nonumber \\  
    &&g_{\mu\nu,\alpha \beta} = O\Bigl(\frac{1}{r^3}\Bigr)  ,\;\;  \ldots .
    \label{aymp1}
\end{eqnarray}
These conditions immediately imply that the Christoffel symbols satisfy
   ${\Gamma^\alpha}_{\mu \nu}=O(r^{-2})$, the extrinsic curvature
$K_{\mu \nu}=O(r^{-2})$ and their derivatives behave as
${\Gamma^\alpha}_{\mu \nu,\beta}=O(r^{-3})$ and
$K_{\mu \nu,\beta}=O(r^{-3})$ etc.\ 
from which the Ricci curvature is $R_{\mu\nu}=O(r^{-3})$, implying
$T_{\mu\nu}=O(r^{-3})$ as well. Again, these are generic conditions, and
the energy-momentum may actually fall to zero more rapidly.

\section{ADM mass}

The ADM mass is defined as a
surface integral at spatial infinity\cite{York1980} (on a Euclidean
sphere there at $r={\rm constant}$)
\begin{eqnarray}
  M^{\text{ADM}} = \frac{1}{16\pi} \int_{\partial\Sigma}
(g_{ij}{}^{,j} - g_{jj,i}) \hat{N}^i dA \;,
  \label{ADM_standard}
\end{eqnarray}
where $\hat{N}^i$
is the unit outward normal to the
spherical boundary at spatial infinity, and $dA$ is an element of
area there.

\section{Equality of ADM mass and \\
Komar energy function in asymptotically-flat
spacetimes}

We now state the main results of this paper.

\begin{widetext}


\noindent
{\bf Theorem for York-lite asymptotic conditions:}\newline
For York-lite asymptotically-flat spacetimes, 
then for the vector field $\xi^\mu = (\partial_t)^\mu + O(r^{-n})$, $n > 0$, 
\begin{eqnarray}
M^{\text{ADM}} &=& E(\xi)
-\frac{1}{8\pi} \int_{\partial\Sigma} G_{\mu\nu}
\hat N^\mu x^\nu dA 
 + \frac{1}{8 \pi} \int_{\partial\Sigma}
\,\Bigl[  \Bigl(( \mathfrak{L }_\xi \,g_{\sigma \beta} )^{;\sigma}
- ( \mathfrak{L }_\xi \; g_{\lambda \sigma} )_{;\beta}\,
g^{\sigma \lambda} \Bigr)
g_{\nu\alpha}
+  ( \mathfrak{L }_\xi \,g_{\nu \alpha} )_{;\beta}
\Bigr] x^\nu \hat N^{[\alpha}\hat T^{\beta]} \, dA,\nonumber \\
\end{eqnarray}
where $G_{\mu\nu}=R_{\mu\nu} -\frac{1}{2} g_{\mu\nu} R$ is the
4-dimensional Einstein tensor and
$\mathfrak{L }_\xi$ denotes the Lie derivative with respect to
the vector field $\xi^\mu$. Note that a function $g(r)=o(r^{-m})$, when
$\vert g(r)\vert \le \epsilon \,r^{-m}$ for {\it every\/} $\epsilon >0$ for 
any sufficiently large $r$.

\vskip 0.1in

When $\xi^\mu = (\partial_t)^\mu + O(r^{-n})$, $n > 0$, is Killing, and
since $\hat N^\mu =(0,\hat N^i)$, then from Eq.~(\ref{TmunuCond})
provided the 4-dimensional Einstein tensor satisfies $G_{ij} =
o(r^{-{3}})$ or equivalently, $T_{ij} = o(r^{-{3}})$, then the above
result straightforwardly reduces to an equality between ADM and Komar
masses.\cite{beig1978,Chrusciel1986}

We now show that an even more elegant result is possible when applying
the Weinberg asymptotic conditions. Recall that the key difference
between the York-lite and Weinberg conditions refers to the action of
temporal derivatives on corrections to the flat spacetime metric at
spatial infinity, as in Eq.~(\ref{aymp1}). In the following theorem we
assume that we may extend this behavior to derivatives on $\xi^\mu$.

\noindent
{\bf Theorem for Weinberg asymptotic conditions:}\newline
For Weinberg asymptotically-flat spacetimes, with the choice $\xi^\mu = {\cal N}\hat T^\mu+o(r^{-1})$,
we find
\begin{equation}
M^{\text{ADM}} = E(\xi) - \frac{1}{8 \pi} \int_{\partial\Sigma} \!\!
{}^{(3)}G_{ij} \hat{N}^i x^j  dA .
\label{TheoremWeinberg}
\end{equation}
provided derivatives on ${\xi^\mu}$ behave as 
$\partial_\nu{\xi^\mu} = \partial_\nu({\cal N}\hat T^\mu)+o(r^{-2})$, 
and similar expressions to higher-order. Here, ${}^{(3)}G_{ij}$ is the
three-dimensional Einstein tensor defined on the hypersurface.

\vskip 0.1in

Note, that when $\xi^\mu = {\cal N}\hat T^\mu$, Komar called
$E({\cal N}\hat T^\mu)$ the generalized Komar energy for
asymptotically-flat dynamical spacetimes,\cite{komar1962}
though he made no claim about its connection to the ADM mass.

\vskip 0.1in

\noindent
{\bf Corollary to `Theorem for Weinberg asymptotic conditions:'}\newline
For Weinberg asymptotically-flat spacetimes, with ${}^{(3)}G_{ij}
= o(r^{-{3}})$,
then the choice $\xi^\mu = {\cal N}\hat T^\mu +o(r^{-1})$ yields
\begin{equation}
M^{\text{ADM}} = E(\xi),
\end{equation}
provided derivatives on ${\xi^\mu}$ behave as
$\partial_\nu{\xi^\mu} = \partial_\nu({\cal N}\hat T^\mu)+o(r^{-2})$, etc.

\vskip 0.1in

\noindent
{\bf Proof of `Theorem for York-lite asymptotic conditions:'}

To connect the ADM mass with the generalized Komar energy, we begin by
transforming the ADM mass into a Komar integral at spatial infinity,
leveraging the asymptotic flatness conditions discussed above. While our
analysis builds closely upon the work of Chru\'{s}ciel
\cite{Chrusciel1986,Chrusciel2010}, his key results were established for
stationary spacetimes possessing a Killing vector. Consequently, we
present a detailed proof, with explicitly stated assumptions, applicable
to asymptotically-flat dynamical spacetimes as mentioned in the statement
of the Theorem.

With the York-lite asymptotic conditions, we may transform
Eq.~(\ref{ADM_standard}) into
\begin{eqnarray}
    M^{\text{ADM}}  &=& \frac{1}{16\pi} \int_{\partial\Sigma}
(g_{ij}{}^{,j} - g_{jj,i}) \hat{N}^i dA \nonumber \\
&=&\frac{1}{16 \pi} \int_{\partial\Sigma} \Bigl( \eta^{i \sigma}
\eta^{j \rho} - \eta^{i \rho} \eta^{j \sigma} \Bigr)
g_{\sigma j ,\rho} \hat N_i dA \nonumber  \\ 
    &=& \frac{-3}{8 \pi} \int_{\partial\Sigma} \frac{1}{3}
\Bigl(  \delta^{[0}_\lambda \delta^{i]}_\mu \delta^j_0 +
\delta^{[i}_\lambda \delta^{j]}_\mu \delta^0_0 +
\delta^{[j}_\lambda \delta^{0]}_\mu \delta^i_0 \Bigr)
\eta^{\lambda \rho} \eta^{\mu \sigma} g_{\sigma j ,\rho}
\hat N_i dA  \nonumber  \\
    &=& \frac{-3}{8 \pi} \int_{\partial\Sigma}
\delta^{[0}_\lambda \delta^i_\mu \delta^{j]}_0 
\eta^{\lambda \rho} \eta^{\mu \sigma} g_{\sigma j ,\rho}
\hat N_i  dA \nonumber  \\
    &=& \frac{3}{8 \pi} \int_{\partial\Sigma}
\delta^{[\beta}_\lambda \delta^\alpha_\mu
\delta^{\gamma]}_\nu \xi^\nu \eta^{\lambda \rho}
\eta^{\mu \sigma} g_{\sigma \gamma ,\rho} \hat N_\alpha\hat T_\beta dA ,
\label{ADMle1}
\end{eqnarray}
where in obtaining line two we recall that latin indices ($i$, $j$, etc)
refer to the spatial components,
$A^{[\alpha\beta\gamma]}\equiv\frac{1}{3!}(A^{\alpha\beta\gamma}
+ \text{anti-symmetrized terms}$).
In the last step we used $\xi^\mu=\delta^\mu_0 + O(r^{-n})$, $n>0$
and $\hat T_\mu=(-{\cal N}, 0, 0, 0)$ with
${\cal N} =1+O(r^{-1})$; this ensures that the index $\beta=0$ and
anti-symmetry among the indices $[\beta,\alpha,\gamma]$ then ensures that 
$\alpha$ and $\gamma$ must be spatial indices. Finally, from the York-lite
asymptotic conditions the only potentially `dangerous' terms could come
from the temporal derivatives $g_{0j,0}=O(r^{-1})$ implying indices
$\sigma=\rho=0$ which in turn require the indices $\mu=\lambda=0$ implying
that such `dangerous' contributions identically vanish.

The expression in Eq.~(\ref{ADMle1}) can be rewritten using properties
of the Levi-Civita tensor and exterior derivatives.\cite{Wald1984}
Since $- 3! \, \delta^{[\beta}_\lambda \delta^\alpha_\mu
\delta^{\gamma]}_\nu = \varepsilon^{\tau\beta\alpha\gamma}
\varepsilon_{\tau\lambda\mu\nu}$,
$\hat N_{[\alpha} \hat T_{\beta]} dA = dS_{\alpha
\beta} = \frac{1}{2} \varepsilon_{\alpha \beta \tau_1 \tau_2}
dx^{\tau_1} \wedge dx^{\tau_2}$ \cite{Chrusciel1986,Chrusciel2010},
Eq.~(\ref{ADMle1}) may be further simplified as
\begin{eqnarray}
    M^{\text{ADM}} &=& \frac{3}{8 \pi} \int_{\partial\Sigma}
\frac{-1}{3!} \varepsilon^{\tau\beta\alpha\gamma}
\varepsilon_{\tau\lambda\mu\nu}\, \xi^\nu \eta^{\lambda \rho}
\eta^{\mu \sigma} g_{\sigma \gamma ,\rho} \Bigl( \frac{1}{2}
\varepsilon_{\alpha \beta \tau_1 \tau_2}
dx^{\tau_1} \wedge dx^{\tau_2} \Bigr)  \nonumber  \\ 
    &=& \frac{1}{8 \pi} \int_{\partial\Sigma}
\frac{-1}{4} \varepsilon_{\tau\lambda\mu\nu}\,
\xi^\nu\eta^{\lambda \rho} \eta^{\mu \sigma} g_{\sigma \gamma ,\rho}
(\varepsilon^{\tau\beta\alpha\gamma}
\varepsilon_{\alpha \beta \tau_1 \tau_2})\, dx^{\tau_1} \wedge dx^{\tau_2} 
    \nonumber  \\ 
    &=&   \frac{1}{8 \pi} \int_{\partial\Sigma}
\frac{-1}{4} \varepsilon_{\tau\lambda\mu\nu}\,
\xi^\nu \eta^{\lambda \rho} \eta^{\mu \sigma}
g_{\sigma \gamma ,\rho} ( 2! 2! )\, \delta^\tau_{[\tau_1}
\delta^\gamma_{\tau_2]}\,  dx^{\tau_1} \wedge dx^{\tau_2}  \nonumber  \\ 
    &=&   \frac{-1}{8 \pi} \int_{\partial\Sigma}
\varepsilon_{\tau\lambda\mu\nu}\, \xi^\nu \eta^{\lambda \rho}
\eta^{\sigma \mu} g_{\sigma \gamma,\rho}\, dx^{\tau} \wedge dx^{\gamma} 
 \nonumber  \\ 
    &=&   \frac{-1}{8 \pi} \int_{\partial\Sigma}
\varepsilon_{\tau\lambda\mu\nu}\, \xi^\nu \eta^{\lambda \rho}
\eta^{\sigma \mu} ( \Gamma_{ \sigma \gamma \rho }
+ \Gamma_{\gamma \sigma \rho } )\, dx^{\tau} \wedge dx^{\gamma}
 \nonumber  \\ 
    &=& \frac{-1}{8 \pi} \int_{\partial\Sigma}
\varepsilon_{\tau\lambda\mu\nu}\, \xi^\nu \eta^{\lambda \rho}
{\Gamma^\mu}_{\gamma \rho }\, dx^{\tau} \wedge dx^{\gamma} ,
\label{ADMle11}
\end{eqnarray}
where we have used $g_{\sigma\gamma, \rho} = \Gamma_{ \sigma \gamma \rho
} + \Gamma_{\gamma \sigma \rho }$ in moving from the fourth to the fifth
line, and $\Gamma_{\gamma \sigma \rho }$ in the fifth line vanishes
because the symmetric indices $\sigma$ and $\rho$ are mapping to an
anti-symmetric tensor. Note, that the indices $\tau$ and $\gamma$ must be
purely spatial from the definition of $dS_{\alpha\beta}$ and $\hat T_\beta$
having only temporal components; this ensures that no $O(r^{-1})$
terms contribute to the Christoffel symbol from the York-lite conditions.

To Further simplify Eq.~(\ref{ADMle11}), we first introduce some
differential tricks we will use. Since $d \sqrt{-g} = \frac{1}{2}
\sqrt{-g} \, g^{\delta\beta} g_{\delta\beta ,\alpha} dx^\alpha$ and
$\varepsilon_{\tau\lambda\mu\nu} = \sqrt{-g}\, [\tau\lambda\mu\nu]$, we
have
\begin{eqnarray}
     d \varepsilon_{\tau\lambda\mu\nu} &=&
\frac{1}{2} \varepsilon_{\tau\lambda\mu\nu} g^{\delta\beta}
g_{\delta\beta , \alpha} dx^\alpha \nonumber  \\ 
    &=& \frac{1}{2} \varepsilon_{\tau\lambda\mu\nu}
g^{\delta \beta} ( \Gamma_{ \delta\beta \alpha} +
\Gamma_{\beta \delta \alpha }) dx^\alpha \nonumber  \\ 
    &=& \varepsilon_{\tau\lambda\mu\nu} {\Gamma^\beta}_{\beta \alpha }
dx^\alpha
    \label{difepsilon}
\end{eqnarray}

Using Leibnitz's rule for the exterior derivative, Eq.~(\ref{ADMle11})
may be simplified as
\begin{eqnarray}
    M^{\text{ADM}} &=& \frac{-1}{8 \pi} \int_{\partial\Sigma}
\varepsilon_{\tau\lambda\mu\nu} \, \xi^\nu \eta^{\lambda \rho}
{\Gamma^\mu}_{\gamma \rho } \, dx^{\tau} \wedge dx^{\gamma}  \nonumber  \\ 
    &=&  \frac{-1}{8 \pi} \int_{\partial\Sigma}
d (\varepsilon_{\tau\lambda\mu\nu}\, \xi^\nu \eta^{\lambda \rho}
{\Gamma^\mu}_{\gamma \rho } x^{\tau} dx^{\gamma} )
-  x^{\tau} d (\varepsilon_{\tau\lambda\mu\nu}\, \xi^\nu \eta^{\lambda \rho}
{\Gamma^\mu}_{\gamma \rho } )  \wedge dx^{\gamma}   \nonumber  \\ 
    &=&  \frac{1}{8 \pi} \int_{\partial\Sigma}  x^{\tau}
d (\varepsilon_{\tau\lambda\mu\nu}\, \xi^\nu \eta^{\lambda \rho}
{\Gamma^\mu}_{\gamma \rho } )  \wedge dx^{\gamma}  \nonumber  \\ 
    &=&  \frac{1}{8 \pi} \int_{\partial\Sigma}  x^{\tau} \eta^{\lambda \rho}
\Bigl( \xi^\nu {\Gamma^\mu}_{\gamma \rho }\, d\varepsilon_{\tau\lambda\mu\nu}
 \wedge dx^{\gamma}
+ \varepsilon_{\tau\lambda\mu\nu} {\Gamma^\mu}_{\gamma \rho }
\, d \xi^\nu \wedge dx^{\gamma}
+ \varepsilon_{\tau\lambda\mu\nu}\, \xi^\nu d {\Gamma^\mu}_{\gamma \rho }
\wedge dx^{\gamma} \Bigr)  \nonumber  \\ 
    &=&  \frac{1}{8 \pi} \int_{\partial\Sigma}  x^{\tau} \eta^{\lambda \rho}
\Bigl( \xi^\nu {\Gamma^\mu}_{\gamma \rho } \,\varepsilon_{\tau\lambda\mu\nu}
{\Gamma^\beta}_{\beta \alpha }\, dx^\alpha \wedge dx^{\gamma}
 + \varepsilon_{\tau\lambda\mu\nu} {\Gamma^\mu}_{\gamma \rho }
\,O\Bigl(\frac{1}{r^{1+n}}\Bigr)^\nu_\alpha \,dx^\alpha \wedge dx^{\gamma}
+ \varepsilon_{\tau\lambda\mu\nu} \,\xi^\nu d {\Gamma^\mu}_{\gamma \rho }
\wedge dx^{\gamma} \Bigr)  \nonumber  \\ 
    &=&  \frac{1}{8 \pi} \int_{\partial\Sigma}  x^{\tau} \eta^{\lambda \rho}
\varepsilon_{\tau\lambda\mu\nu} \, \xi^\nu {\Gamma^\mu}_{\gamma \rho ,\alpha}
\,dx^\alpha \wedge dx^{\gamma}
= \frac{1}{8 \pi} \int_{\partial\Sigma}  x^{\tau} g^{\lambda \rho}
\varepsilon_{\tau\lambda\mu\nu}\,\xi^\nu {\Gamma^\mu}_{\rho [\gamma ,\alpha]}
\,dx^\alpha \wedge dx^{\gamma} ,
\label{ADMle13}
\end{eqnarray}
where Stokes' theorem and the boundary of a boundary is an empty set
are used in the second line, and Eq.~(\ref{difepsilon}). In going
from the fourth to fifth line only spatial derivatives to $d\xi^\nu$
can contribute as $dx^\alpha$ is tangent to a boundary at constant $t$.
In the fifth line, the indices $\alpha$ and $\gamma$ are both spatial since
they are tangent to the boundary and hence from the asymptotic
conditions in Eq.~(\ref{ChrisCond}) the first term vanishes.

Then again since the indices $\alpha$ and $\gamma$ are purely spatial,
we immediately have
${\Gamma^\mu}_{\rho [\gamma ,\alpha]} = -\frac{1}{2}
{R^\mu}_{\rho \gamma \alpha} + O(r^{-4})$. Further, since
$dx^\alpha \wedge dx^{\gamma} = - \frac{1}{2}
\varepsilon^{\alpha \gamma \tau_1 \tau_2} dS_{\tau_1 \tau_2}$,
Eq.~(\ref{ADMle13}) may be simplified as
\begin{eqnarray}
    M^{\text{ADM}} &=& \frac{1}{8 \pi} \int_{\partial\Sigma}
\varepsilon_{\tau\lambda\mu\nu} \, x^{\tau} \xi^\nu g^{\lambda \rho}
\Bigl(-\frac{1}{2} {R^\mu}_{\rho \gamma \alpha} + O\Bigl(\frac{1}{r^4}\Bigr)
\Bigr) dx^\alpha \wedge dx^{\gamma}   \nonumber  \\ 
    &=&  \frac{1}{16 \pi} \int_{\partial\Sigma}
\varepsilon_{\mu\lambda\nu\tau} \,\xi^\nu x^{\tau}
{R^{\mu \lambda}}_{ \alpha \gamma}\, dx^\alpha \wedge dx^{\gamma} 
= \frac{1}{16 \pi} \int_{\partial\Sigma} \varepsilon_{\mu\lambda\nu\tau}
\,\xi^\nu x^{\tau} {R^{\mu \lambda}}_{ \alpha \gamma}
(- \frac{1}{2} \varepsilon^{\alpha \gamma \tau_1 \tau_2}
dS_{\tau_1 \tau_2})  \nonumber  \\ 
    &=&   \frac{-1}{32 \pi} \int_{\partial\Sigma}
\varepsilon_{\mu\lambda\nu\tau} \,\varepsilon^{\alpha \gamma \tau_1 \tau_2}
\xi^\nu x^{\tau} {R^{\mu \lambda}}_{ \alpha \gamma}\, dS_{\tau_1 \tau_2}
= \frac{-1}{32 \pi} \int_{\partial\Sigma} (- 4! \delta^\alpha_{[\mu}
\delta^\gamma_\lambda \delta^{\tau_1}_\nu \delta^{\tau_2}_{\tau ]})
\, \xi^\nu x^{\tau} {R^{\mu \lambda}}_{ \alpha \gamma}\, dS_{\tau_1 \tau_2}  .
\label{ADMle14}
\end{eqnarray}
As $ \delta^\alpha_{[\mu} \delta^\gamma_\lambda \delta^{\tau_1}_\nu
\delta^{\tau_2}_{\tau ]}$ may be expanded as
\begin{eqnarray}
    \delta^\alpha_{[\mu} \delta^\gamma_\lambda \delta^{\tau_1}_\nu
\delta^{\tau_2}_{\tau ]} &=& \frac{1}{3!}
\biggl( \delta^\alpha_{[\mu} \delta^\gamma_{\lambda]}
\delta^{\tau_1}_{[\nu} \delta^{\tau_2}_{\tau ]}
- \delta^\alpha_{[\mu} \delta^\gamma_{\nu]} \delta^{\tau_1}_{[\lambda}
\delta^{\tau_2}_{\tau ]}
+ \delta^\alpha_{[\mu} \delta^\gamma_{\tau]} \delta^{\tau_1}_{[\lambda}
\delta^{\tau_2}_{\nu ]}
+ \delta^\alpha_{[\lambda} \delta^\gamma_{\nu]} \delta^{\tau_1}_{[\mu}
\delta^{\tau_2}_{\tau ]}
+ \delta^\alpha_{[\lambda} \delta^\gamma_{\tau]} \delta^{\tau_1}_{[\nu}
\delta^{\tau_2}_{\mu ]}
+ \delta^\alpha_{[\tau} \delta^\gamma_{\nu]} \delta^{\tau_1}_{[\lambda}
\delta^{\tau_2}_{\mu ]} \biggr)  \nonumber  \\ 
    &=& \frac{1}{3!} \biggl( \delta^{[\alpha}_\mu \delta^{\gamma]}_\lambda
\delta^{[\tau_1}_\nu \delta^{\tau_2]}_\tau 
- \delta^{[\alpha}_\mu \delta^{\gamma]}_\nu \delta^{[\tau_1}_\lambda
\delta^{\tau_2]}_\tau 
+ \delta^{[\alpha}_\mu \delta^{\gamma]}_\tau \delta^{[\tau_1}_\lambda
\delta^{\tau_2]}_\nu 
+ \delta^{[\alpha}_\lambda \delta^{\gamma]}_\nu \delta^{[\tau_1}_\mu
\delta^{\tau_2]}_\tau  + \delta^{[\alpha}_\lambda \delta^{\gamma]}_\tau
\delta^{[\tau_1}_\nu \delta^{\tau_2]}_\mu + \delta^{[\alpha}_\tau
\delta^{\gamma]}_\nu \delta^{[\tau_1}_\lambda \delta^{\tau_2]}_\mu \biggr),
 \nonumber
    \label{delta}
\end{eqnarray}
We see that Eq.~(\ref{ADMle14}) becomes
\begin{eqnarray}
   M^{\text{ADM}} &=& \frac{1}{8 \pi} \int_{\partial\Sigma}
3! (\delta^\alpha_{[\mu} \delta^\gamma_\lambda \delta^{\tau_1}_\nu
\delta^{\tau_2}_{\tau ]})\, \xi^\nu x^{\tau}
{R^{\mu \lambda}}_{ \alpha \gamma}\, dS_{\tau_1 \tau_2} \nonumber  \\ 
    &=&  \frac{1}{8 \pi} \int_{\partial\Sigma}  \biggl( \delta^{[\alpha}_\mu
\delta^{\gamma]}_\lambda \delta^{[\tau_1}_\nu \delta^{\tau_2]}_\tau 
- \delta^{[\alpha}_\mu \delta^{\gamma]}_\nu \delta^{[\tau_1}_\lambda
\delta^{\tau_2]}_\tau 
+ \delta^{[\alpha}_\mu \delta^{\gamma]}_\tau \delta^{[\tau_1}_\lambda
\delta^{\tau_2]}_\nu
+ \delta^{[\alpha}_\lambda \delta^{\gamma]}_\nu \delta^{[\tau_1}_\mu
\delta^{\tau_2]}_\tau  \nonumber  \\ 
    &&+ \delta^{[\alpha}_\lambda \delta^{\gamma]}_\tau \delta^{[\tau_1}_\nu
\delta^{\tau_2]}_\mu
+ \delta^{[\alpha}_\tau \delta^{\gamma]}_\nu \delta^{[\tau_1}_\lambda
\delta^{\tau_2]}_\mu \biggr)
\xi^\nu x^{\tau} {R^{\mu \lambda}}_{ \alpha \gamma}
\, dS_{\tau_1 \tau_2} \nonumber  \\ 
    &=&  \frac{1}{8 \pi} \int_{\partial\Sigma}  \xi^\nu x^{\tau}
\biggl( {R^{\mu \lambda}}_{ \mu \lambda} \,dS_{\nu \tau}  
-  {R^{\mu \lambda}}_{ \mu \nu} \,dS_{\lambda \tau} 
+  {R^{\mu \lambda}}_{ \mu \tau} \,dS_{\lambda \nu} 
+  {R^{\mu \lambda}}_{ \lambda \nu} \,dS_{\mu \tau} 
+  {R^{\mu \lambda}}_{ \lambda \tau} \,dS_{\nu \mu} 
+  {R^{\mu \lambda}}_{ \tau \nu} \,dS_{\lambda \mu}  \biggr)  \nonumber  \\ 
    &=&  \frac{1}{8 \pi} \int_{\partial\Sigma} 
\xi^\nu x^{\tau} \Bigl( R  \,dS_{\nu \tau} - 2{R^\mu}_\nu \,dS_{\mu \tau}
+  2{R^\mu}_\tau\,dS_{\mu \nu}
+  {R^{\mu \lambda}}_{ \tau \nu}\, dS_{\lambda \mu}  \Bigr) .
\label{ADMle15}
\end{eqnarray}

Although this expression may be simplified by requiring
$T_{\mu\nu}= o(r^{-3})$, as Chru\'{s}ciel assumes in deriving his
ADM formula,\cite{Chrusciel1986,Chrusciel2010} we prefer to analyze
the asymptotically-flat conditions in more detail in order to achieve
a weaker assumption. Recalling that
$dS_{\lambda \mu} = \hat N_{[\lambda} \hat T_{\mu]} dA$
we now further simplify Eq.~(\ref{ADMle15}) as
\begin{eqnarray}
M^{\text{ADM}}
    &=& \frac{1}{8 \pi} \int_{\partial\Sigma}  \xi^\nu x^{\tau}
    \Bigl( R \hat N_{[\nu} \hat T_{\tau]}
    - 2{R^\mu}_\nu \hat N_{[\mu} \hat T_{\tau]}
    +  2{R^\mu}_\tau  \hat N_{[\mu} \hat T_{\nu]} \Bigr) dA
    + \frac{1}{8\pi} \int_{\partial\Sigma}
    \xi^\mu x^\nu R_{\mu\nu\alpha\beta}
    \hat N^\alpha \hat T^\beta dA  \nonumber \\ 
    &=&  \frac{1}{16 \pi} \int_{\partial\Sigma} 
\xi^\nu x^{\tau} \Bigl( - R \hat T_\nu \hat N_\tau +
2{R^\mu}_\nu \hat T_\mu \hat N_\tau
+  2{R^\mu}_\tau  \hat N_\mu \hat T_\nu \Bigr) dA
+ \frac{1}{8\pi} \int_{\partial\Sigma}
\xi^\mu x^\nu R_{\mu\nu\alpha\beta} \hat N^\alpha \hat T^\beta dA 
\nonumber  \\ 
    &=&  \frac{1}{16 \pi} \int_{\partial\Sigma}
\Bigl( R \hat N_\tau x^{\tau}
+ 2 R_{\mu\nu} \hat T^\mu \xi^\nu \hat N_\tau x^{\tau}
-  2 R_{\mu\tau}  \hat N^\mu \hat x^{\tau} \Bigr) dA
+ \frac{1}{8\pi} \int_{\partial\Sigma} \xi^\mu x^\nu
R_{\mu\nu\alpha\beta} \hat N^\alpha \hat T^\beta dA , \nonumber \\
&=& -\frac{1}{8\pi} \int_{\partial\Sigma} (R_{\mu\nu} - \frac{1}{2} R g_{\mu\nu} ) \hat N^\mu x^\nu dA
+ \frac{1}{8\pi} \int_{\partial\Sigma} 
 \xi^\mu x^\nu R_{\mu\nu\alpha\beta} \hat N^\alpha \hat T^\beta dA 
+ \frac{1}{8\pi} \int_{\partial\Sigma} 
 R_{\mu\nu} \,\xi^\mu \hat T^\nu \hat N_\tau x^{\tau} dA .
 \label{ourADM}
\end{eqnarray}
In moving from the first to the second line, we use the following
\begin{eqnarray}
R \,\xi^\nu \hat N_\nu \,x^\tau \hat T_\tau
&=& O\Bigl(\frac{1}{r^2}\Bigr) O\Bigl(\frac{1}{r^n}\Bigr) 
\Bigl[ -t +O\Bigl(\frac{1}{r}\Bigr) \Bigr]
= -t\, O\Bigl(\frac{1}{r^{2+n}}\Bigr) = O\Bigl(\frac{1}{r^{2+n}}\Bigr)
\nonumber\\
R_{\mu\nu} \hat N^\mu \xi^\nu x^\tau \hat T_\tau &=& 
\bigl( R_{i0} \xi^0 + R_{ij}  \xi^j\bigr) \hat N^i
\Bigl[ -t +O\Bigl(\frac{1}{r}\Bigr) \Bigr]
= -t \, \Bigl[ O\Bigl(\frac{1}{r^3}\Bigr) 
+O\Bigl(\frac{1}{r^2}\Bigr) O\Bigl(\frac{1}{r^n}\Bigr) \Bigr] \hat N^i
=  O\Bigl(\frac{1}{r^{2+n}}\Bigr)
\nonumber\\
{R^\mu}_\tau \, \hat T_\mu \,x^\tau  \xi^\nu \hat N_\nu  &=&
\bigl( t\, {R^0}_0  + x^i {R^0}_i  \bigr) \, \hat T_0 \,
O\Bigl(\frac{1}{r^n}\Bigr)
= \Bigl[ t \, O\Bigl(\frac{1}{r^{2+n}}\Bigr)
+ x^i O\Bigl(\frac{1}{r^{3+n}}\Bigr)\Bigr]
= O\Bigl(\frac{1}{r^{2+n}}\Bigr),
\label{RmunuCond}
\end{eqnarray}
which follow from the York-lite asymptotic conditions
$R_{00}=O(r^{-2})$ and $R_{ij}=O(r^{-2})$ although $R_{0i}=O(r^{-3})$
by Eq.~(\ref{RicciCond}), and from
with $\xi^\mu \hat N_\mu = O(r^{-n})$, $n>0$, and
$x^\beta \hat T_\beta = -t +O(r^{-1})$
where $t=\text{constant}$ on the hypersurface $\Sigma$. 
In the final step of Eq.~(\ref{ourADM}) we use the Einstein field
equations on the first and third terms.

Before we continue our transformation of the ADM mass, let us first
prove a lemma that we will use soon. 

\noindent
{\bf Lemma 1:}\newline
For any vector field $\xi^\mu$ and coordinates $x^\mu$
\begin{equation}
2 \xi^{[\alpha;\beta]}
= - 3( \xi^{[\beta;\alpha} x^{\nu]} )_{;\nu}
+ {\xi^{[\beta;\alpha]}}_{;\nu} x^\nu 
+ {\xi^{[\nu ; \beta]}}_{;\nu} x^\alpha
+ {\xi^{[\alpha;\nu]}}_{;\nu} {x^\beta}.
\end{equation}

\noindent
{\bf Proof of Lemma 1:}\newline
Coordinates are scalar functions, so
$\delta^\beta_\nu={x^\beta}_{,\nu}={x^\beta}_{;\nu}$ thus
$3 \xi^{[\beta;\alpha} x^{\nu]} = \xi^{[\beta;\alpha]} x^\nu
+ \xi^{[\alpha ; \nu]} x^\beta + \xi^{[\nu ; \beta]} x^\alpha$, and hence
\begin{eqnarray}
   \xi^{[\alpha;\beta]} &=& \xi^{[\alpha;\nu]} \delta^\beta_\nu
= \xi^{[\alpha;\nu]} {x^\beta}_{;\nu}
= (\xi^{[\alpha;\nu]} x^\beta)_{;\nu}
- {\xi^{[\alpha;\nu]}}_{;\nu} {x^\beta} 
   =  (3 \xi^{[\beta;\alpha} x^{\nu]} - \xi^{[\beta;\alpha]} x^\nu
- \xi^{[\nu ; \beta]} x^\alpha )_{;\nu}
- {\xi^{[\alpha;\nu]}}_{;\nu} {x^\beta}  \nonumber \\ 
   &=& 3( \xi^{[\beta;\alpha} x^{\nu]} )_{;\nu}
- {\xi^{[\beta;\alpha]}}_{;\nu} x^\nu
- \xi^{[\beta;\alpha]} {x^\nu}_{;\nu} 
- {\xi^{[\nu ; \beta]}}_{;\nu} x^\alpha 
- \xi^{[\nu ; \beta]} {x^\alpha }_{;\nu} 
- {\xi^{[\alpha;\nu]}}_{;\nu} {x^\beta}  \nonumber \\ 
   &=& 3( \xi^{[\beta;\alpha} x^{\nu]} )_{;\nu}
- {\xi^{[\beta;\alpha]}}_{;\nu} x^\nu - 4 \xi^{[\beta;\alpha]} 
- {\xi^{[\nu ; \beta]}}_{;\nu} x^\alpha  - \xi^{[\alpha ; \beta]}
- {\xi^{[\alpha;\nu]}}_{;\nu} {x^\beta} \nonumber \\ 
   &=& 3( \xi^{[\beta;\alpha} x^{\nu]} )_{;\nu}
- {\xi^{[\beta;\alpha]}}_{;\nu} x^\nu + 3 \xi^{[\alpha;\beta]} 
- {\xi^{[\nu ; \beta]}}_{;\nu} x^\alpha
- {\xi^{[\alpha;\nu]}}_{;\nu} {x^\beta} \;,
\end{eqnarray}
or equivalently, we obtain the claim of the Lemma that
\begin{eqnarray}
   2 \xi^{[\alpha;\beta]} = -3( \xi^{[\beta;\alpha} x^{\nu]} )_{;\nu}
+ {\xi^{[\beta;\alpha]}}_{;\nu} x^\nu 
+ {\xi^{[\nu ; \beta]}}_{;\nu} x^\alpha
+ {\xi^{[\alpha;\nu]}}_{;\nu} {x^\beta} \;.
\qquad \qquad \qed
 \label{Lemma3}
\end{eqnarray}


Consequently:
\begin{eqnarray}
   {\xi^{[\beta;\alpha]}}_{;\nu} x^\nu = 2 \xi^{[\alpha;\beta]}
+ 3( \xi^{[\beta;\alpha} x^{\nu]} )_{;\nu}
- {\xi^{[\nu ; \beta]}}_{;\nu} x^\alpha
- {\xi^{[\alpha;\nu]}}_{;\nu} {x^\beta} \;.
 \label{Lemma4}
\end{eqnarray}

Recall that permuting the order of a pair of covariant derivatives
acting on an arbitrary 4-vector $\xi^\mu$ may be expressed in terms of the
Riemann curvature tensor as \cite{poisson2004}
${\xi^\mu}_{;\alpha \beta} - {\xi^\mu}_{;\beta \alpha } 
    = - {R^\mu}_{ \nu \alpha \beta} \xi^\nu$.
Contracting the indices $\mu$ and $\alpha$ reduces this to
an expression in terms of the Ricci tensor
${\xi^\mu}_{;\mu \beta} - {\xi^\mu}_{;\beta \mu } 
    = - R_{ \nu \beta} \xi^\nu$.
Consequently, for an arbitrary $\xi^\mu$, we may write
$J_\beta(\xi)={\xi_{[\mu ; \beta]}}^{;\mu}  
 = R_{ \mu \beta} \xi^\mu + {\xi^\mu}_{;\mu \beta}
- {\xi_{\{\mu ; \beta\}}} ^{; \mu}$.
\def\ahss{
{\color{red}
From the Ricci identity $R_{\mu \nu \alpha \beta} \xi^\mu =
\xi_{\nu;\alpha\beta} - \xi_{\nu;\beta \alpha} $, 
and recalling 
$\mathfrak{L }_\xi g_{\mu \nu}=\xi_{\mu;\nu}+\xi_{\nu;\mu}$
and $R_{\mu \alpha\beta\gamma}+R_{\mu \beta\gamma\alpha}
+R_{\mu \gamma\alpha\beta}=0$
we have
$R_{\mu\nu\alpha\beta} \xi^\mu = \xi_{[\beta;\alpha];\nu} + \frac{1}{2}
( \mathfrak{L }_\xi g_{\nu \alpha} )_{;\beta} - \frac{1}{2} (\mathfrak{L
}_\xi g_{\beta \nu} )_{;\alpha} $ for any vector field $\xi^\mu$. Then
by contracting $\nu$ and $\beta$, we find
}}
Rewriting this in terms of Lie derivatives we then find
\begin{eqnarray}
    {\xi_{[\nu;\alpha]}}^{;\nu} = R_{\mu \alpha} \xi^\mu
-\frac{1}{2} ( \mathfrak{L }_\xi g_{\nu \alpha} )^{;\nu}
+ \frac{1}{2} ( \mathfrak{L }_\xi g_{\beta \nu} )_{;\alpha} g^{\nu \beta}. 
\label{Rmn5}
\end{eqnarray}

The second integral in
Eq.~(\ref{ourADM}) may now be written as
\begin{eqnarray}
    &&\frac{1}{8\pi} \int_{\partial\Sigma}
\xi^\mu x^\nu R_{\mu\nu\alpha\beta}
\hat N^\alpha \hat T^\beta dA \nonumber \\ 
   &=&\frac{1}{8 \pi} \int_{\partial\Sigma} 
\xi_{[\beta;\alpha];\nu} x^\nu \hat N^\alpha \hat T^\beta dA 
+\frac{1}{16 \pi} \int_{\partial\Sigma} 
x^\nu ( ( \mathfrak{L }_\xi g_{\nu \alpha} )_{;\beta}
- ( \mathfrak{L }_\xi g_{\beta \nu} )_{;\alpha} )
\hat N^\alpha\hat T^\beta dA \nonumber \\ 
   &=&  \frac{1}{8 \pi} \int_{\partial\Sigma} 
\Bigl( 2 \xi^{[\alpha;\beta]}
+ 3( \xi^{[\beta;\alpha} x^{\nu]} )_{;\nu}
- {\xi^{[\nu ; \beta]}}_{;\nu} x^\alpha
- {\xi^{[\alpha;\nu]}}_{;\nu} {x^\beta} \Bigr)
\hat N_\alpha \hat T_\beta dA 
+\frac{1}{16 \pi} \int_{\partial\Sigma} 
x^\nu ( ( \mathfrak{L }_\xi g_{\nu \alpha} )_{;\beta}
- ( \mathfrak{L }_\xi g_{\beta \nu} )_{;\alpha} )
\hat N^\alpha\hat T^\beta dA  \nonumber \\ 
   &=& E(\xi) + \frac{1}{8 \pi} \int_{\partial\Sigma} 
\Bigl( - {\xi^{[\nu ; \beta]}}_{;\nu} x^\alpha
- {\xi^{[\alpha;\nu]}}_{;\nu} {x^\beta} \Bigr)
\hat N_\alpha \hat T_\beta dA 
+\frac{1}{16 \pi} \int_{\partial\Sigma} x^\nu ( ( \mathfrak{L }_\xi
g_{\nu \alpha} )_{;\beta}
- ( \mathfrak{L }_\xi g_{\beta \nu} )_{;\alpha} )
\hat N^\alpha\hat T^\beta dA  \nonumber \\ 
   &=& E(\xi) + \frac{1}{16 \pi} \int_{\partial\Sigma}
\biggl[ 2 \Bigl(  R_{\mu \alpha} \xi^\mu
\hat N^\alpha \hat T_\beta x^\beta
- R_{\mu \beta} \xi^\mu \hat T^\beta \hat N_\alpha x^\alpha \Bigr)
+ \Bigl(( \mathfrak{L }_\xi g_{\nu \beta} )^{;\nu}
- ( \mathfrak{L }_\xi g_{\lambda \nu} )_{;\beta} g^{\nu \lambda} \Bigr)
x^\alpha \hat N_\alpha \hat T^\beta \nonumber \\ 
   &&- \Bigl( ( \mathfrak{L }_\xi g_{\nu \alpha} )^{;\nu}
- ( \mathfrak{L }_\xi g_{\lambda \nu} )_{;\alpha} g^{\nu \lambda} \Bigr)
{x^\beta} \hat N^\alpha \hat T_\beta 
+ \Bigl( ( \mathfrak{L }_\xi g_{\nu \alpha} )_{;\beta}
- ( \mathfrak{L }_\xi g_{\beta \nu} )_{;\alpha} \Bigr)
x^\nu \hat N^\alpha\hat T^\beta  \biggr] dA   \nonumber \\ 
   &=& E(\xi) - \frac{1}{8 \pi} \int_{\partial\Sigma}
\!\! R_{\mu \beta} \,\xi^\mu \hat T^\beta \hat N_\alpha x^\alpha dA
+  \frac{1}{8 \pi} \int_{\partial\Sigma}
\Bigl[\Bigl(( \mathfrak{L }_\xi g_{\nu \beta} )^{;\nu}
- ( \mathfrak{L }_\xi g_{\lambda \nu} )_{;\beta} g^{\nu \lambda} \Bigr)
x_\alpha 
+ (\mathfrak{L }_\xi g_{\nu \alpha} )_{;\beta} \, x^\nu
\Bigr] \hat N^{[\alpha}\hat T^{\beta]}   \, dA . 
 \label{ADMKomar}
\end{eqnarray}
Here, in moving from the second to the third line of
Eq.~(\ref{ADMKomar}) we use Eq.~(\ref{Lemma4}). Next, the second term in
the third line  of Eq.~(\ref{ADMKomar}) vanishes because we may use Stokes'
theorem and relying on the fact that the boundary of
a boundary is empty, and we have also used Eq.~(\ref{Rmn5}) twice in the
fourth line to obtain the fifth line. To go from the fifth line
to the final result of Eq.~(\ref{ADMKomar}), we must kill-off the first
term under the integral using the asymptotic behavior found in
Eq.~(\ref{RmunuCond}), finally obtaining the result by moving the
anti-symmetry back into the measure
$dS^{\alpha\beta}=\hat N^{[\alpha}\hat T^{\beta]}   \, dA$.

\def\xxxK{
In addition, we must kill-off the second pair
of Lie derivatives. To see how this is possible, consider
\begin{equation}
\mathfrak{L}_\xi \,g_{\mu\nu}
\equiv g_{\mu\nu,\tau} \,\xi^\tau + \xi^\tau{}_{,\mu}\, g_{\tau\nu}
+ \xi^\tau{}_{,\nu}\, g_{\mu\tau}
=g_{\mu\nu,0}  + g_{\mu\nu,k} \,\xi^k
+ {\xi^k}_{,\mu}\, g_{k\nu} + {\xi^k}_{,\nu}\, g_{\mu k} ,
\end{equation}
since $\xi^0=1$, and hence
\begin{eqnarray}
\mathfrak{L}_\xi \,g_{00}&=& 
g_{00,0} + g_{00,k} \,\xi^k
+ {\xi^k}_{,0}\, g_{k0} + {\xi^k}_{,0}\, g_{0k}
=g_{00,0} + O\Bigl(\frac{1}{r^{1+n}}\Bigr)
\nonumber \\
\mathfrak{L}_\xi \,g_{i0}&=& 
g_{i0,0} + g_{i0,k} \,\xi^k
+ {\xi^k}_{,i}\, g_{k0} + {\xi^k}_{,0}\, g_{ik} 
= {\xi^k}_{,0}\, g_{ik} + g_{i0,0} + O\Bigl(\frac{1}{r^{2+n}}\Bigr)
\nonumber \\
\mathfrak{L}_\xi \,g_{ij}&=& 
g_{ij,0} + g_{ij,k} \,\xi^k
+ {\xi^k}_{,i}\, g_{kj} + {\xi^k}_{,j}\, g_{ik} 
= O\Bigl(\frac{1}{r^{1+n}}\Bigr)
\end{eqnarray}
where in the final expression we rely on $g_{ij,0}=O(r^{-2})$ from
the York-lite asymptotic conditions found in Eq.~(\ref{aymp0}).

 and asymptotics of second pair of Lie derivatives
are shown to satisfy
\begin{eqnarray}
\Bigl( ( \mathfrak{L }_\xi g_{\nu \alpha} )^{;\nu}
- ( \mathfrak{L }_\xi g_{\lambda \nu} )_{;\alpha} g^{\nu \lambda} \Bigr)
 \hat N^\alpha x^\beta \hat T_\beta 
\end{eqnarray}

Finally, we have applied $x^\beta \hat T_\beta= -t +
O(r^{-1})$ in the last step of Eq.~(\ref{ADMKomar}) and hence the terms
containing $x^\beta \hat T_\beta$ vanish for any $n>0$ and finite $t$.
}

Inserting Eq.~(\ref{ADMKomar}) back into Eq.~(\ref{ourADM}) and
after cancellation of the $R_{\mu\nu}$ term in Eq.~(\ref{ourADM}) yields
\begin{eqnarray}
M^{\text{ADM}} &=&  E(\xi)
-\frac{1}{8\pi} \int_{\partial\Sigma} (R_{\mu\nu}
- \frac{1}{2} R g_{\mu\nu} ) \hat N^\mu x^\nu dA \nonumber \\
&& +\frac{1}{8 \pi} \int_{\partial\Sigma}
\,\Bigl[  \Bigl(( \mathfrak{L }_\xi \,g_{\sigma \beta} )^{;\sigma}
- ( \mathfrak{L }_\xi \; g_{\lambda \sigma} )_{;\beta}\,
g^{\sigma \lambda} \Bigr)
g_{\nu\alpha}
+  ( \mathfrak{L }_\xi \,g_{\nu \alpha} )_{;\beta}
\Bigr] x^\nu \hat N^{[\alpha}\hat T^{\beta]} .
\label{Result1}
\end{eqnarray}

This completes the theorem's proof.
\qed

\vskip 0.1in

\noindent
{\bf Proof of `Theorem for Weinberg asymptotic conditions:'}

Firstly we note that Weinberg's asymptotically-flat conditions fully
satisfy the York-lite conditions as well. Next, as ${\cal N}\,\hat T^\mu
=(1,-\beta^i)$ and by Weinberg's conditions $\beta^i={\cal N}^2
g^{0i}=O(r^{-1})$, we see that $\xi^\mu = {\cal N}\,\hat
T^\mu+o(r^{-1})$ is encompassed by $(\partial_t)^\mu +O(r^{-n})$, $n>0$.
We will also assume that derivatives to the asymptotic corrections to
$\xi^\mu$ exhibit a behavior analogous to that of derivatives of the
asymptotic metric. Specifically, that they satisfy ${\xi^\mu}_{,\nu} =
({\cal N}\hat T^\mu)_{,\nu} +o(r^{-2})$, and similarly for higher-order
derivatives. As all the assumptions necessary to invoke our Theorem for
the York-lite asymptotic conditions, we shall use Eq.~(\ref{Result1}) as
our starting point here.

From the Weinberg's asymptotic conditions, Eq.~(\ref{aymp1}),
the Lie derivative of the metric with respect to $\xi^\mu$ can
be approximated at large $r$ as $\mathfrak{L}_\xi g_{\mu\nu}=O(r^{-2})$,
or, in more detail
\begin{equation}
    \mathfrak{L}_\xi g_{\mu\nu} = g_{\mu\nu,\tau} \xi^\tau
+ \xi^\tau{}_{,\mu} g_{\tau\nu} + \xi^\tau{}_{,\nu} g_{\mu\tau}
=  g_{\mu\nu,0} - \beta^k{}_{,\mu} g_{k\nu} - \beta^k{}_{,\nu} g_{\mu k}
+ o\Bigl(\frac{1}{r^{2}}\Bigr).
    \label{aymp4}
\end{equation}
As the Christoffel symbols are all $O(r^{-2})$, we see that
$(\mathfrak{L}_\xi g_{\mu\nu})_{;\alpha}
=(\mathfrak{L}_\xi g_{\mu\nu})_{,\alpha}+O(r^{-4})$, with the $O(r^{-4})$
terms being too small to contribute. Thus, the behavior of
Eq.~(\ref{aymp1}) and Eq.~(\ref{aymp4}) applied to final integral
in Eq.~(\ref{Result1}) yield
\begin{eqnarray}
    &&\frac{1}{16 \pi} \int_{\partial\Sigma} 
\Bigl[ \Bigl(( \mathfrak{L }_\xi g_{\sigma \beta} )^{,\sigma}
- ( \mathfrak{L }_\xi g_{\lambda \sigma})_{,\beta} g^{\sigma\lambda} \Bigr)
x^\alpha \hat N_\alpha \hat T^\beta
+ \Bigl( ( \mathfrak{L }_\xi g_{\nu \alpha} )_{,\beta}
- ( \mathfrak{L }_\xi g_{\beta \nu} )_{,\alpha} \Bigr)
x^\nu \hat N^\alpha\hat T^\beta
+ O\Bigl(\frac{1}{r^3}\Bigr) \Bigr] dA  
\nonumber \\ 
    &=&\frac{1}{16 \pi} \int_{\partial\Sigma} 
\Bigl[ \Bigl( {g_{\sigma \beta,0}} ^{,\sigma}
- {{\beta^k}_{,\sigma}}^{,\sigma} g_{k \beta} - {\beta^k}_{,\beta k}
- g_{\lambda \sigma,0\beta} \,g^{\sigma \lambda} + {\beta^k}_{, k \beta}
+ {\beta^k}_{, k \beta}
+ o\Bigl(\frac{1}{r^{3}}\Bigr)  \Bigr)
\hat T^\beta x^\alpha \hat N_\alpha  \nonumber \\ 
    &&+ \Bigl( g_{\nu \alpha , 0 \beta}
- {\beta^k}_{,\nu \beta} \, g_{\alpha k}
- {\beta^k}_{, \alpha \beta}\, g_{\nu k}
- g_{\beta \nu ,0 \alpha} + {\beta^k}_{, \beta \alpha}\, g_{\nu k}
+{\beta^k}_{,\nu \alpha} \,g_{\beta k}
+ o\Bigl(\frac{1}{r^{3}}\Bigr) \Bigr)
x^\nu \hat N^\alpha\hat T^\beta \Bigr] dA  .
    \label{flat1}
\end{eqnarray}
Since $\hat{N}^\mu = (0, \hat{N}^i)$ and $x^0=t$ is constant on
$\partial\Sigma$, the dominant contributions in Eq.~(\ref{flat1}) should
be those contracted with the spatial position vector $x^i$ and the
normal vector $\hat{N}^j$. Therefore, Eq.~(\ref{flat1}) may be
further simplified into
\begin{eqnarray}
    &=&\frac{1}{16 \pi} \int_{\partial\Sigma} 
\Bigl[ \Bigl( {g_{l 0,0}} ^{,l}
- {{\beta^k}_{,\sigma}}^{,\sigma} g_{k 0} - {\beta^k}_{,0 k}
- g_{l k,00}\, g^{k l} + {\beta^k}_{, k 0}
+ {\beta^k}_{, k 0}
+ o\Bigl(\frac{1}{r^{3}}\Bigr)  \Bigr)
 x^i \hat N_i  \nonumber \\ 
    &&+ \Bigl( g_{ij , 0 0}
- {\beta^k}_{,i0}\, g_{j k}
- {\beta^k}_{, j0}\, g_{i k}
- g_{0i ,0 j} + {\beta^k}_{, 0j}\, g_{i k}
+{\beta^k}_{,ij}\, g_{0 k}
+ o\Bigl(\frac{1}{r^{3}}\Bigr) \Bigr)
x^i \hat N^j \Bigr] dA \nonumber \\ 
    &=&\frac{1}{16 \pi} \int_{\partial\Sigma} 
\Bigl[ \Bigl( \bigl( {\beta_{l,0}} ^{,l} 
- g_{l k,00}\, g^{k l} 
+ {\beta^k}_{, k 0} \bigr) \gamma_{ij}  
 + \bigl( g_{ij , 0 0} - \beta_{j,i0} - \beta_{i, j0}  \bigr) \Bigr)
x^i \hat N^j + o\Bigl(\frac{1}{r^{2}} \Bigr) \Bigr] dA \nonumber \\ 
    &=&-\frac{1}{8 \pi} \int_{\partial\Sigma} 
\Bigl( \bigl( K_{ij} - K \gamma_{ij} \bigr)_{,0}
x^i \hat N^j + o\Bigl(\frac{1}{r^{2}} \Bigr) \Bigr) dA ,
    \label{flat2}
\end{eqnarray}
where, to obtain the first line we used the fact that
${g_{00,0}}^{,0}-g_{00,00}\, g^{00}=O(r^{-4})$, and
in the final step, the terms in parentheses have been identified
as the time derivative of the extrinsic curvature, $(K_{ij})_{,0}$, and
its trace to $O(r^{-3})$, noting that $g^{kl}=\gamma^{kl}+O(r^{-2})$.

\def \old{
At large distances, the position vector can be approximated
as $x^i = r \hat{N}^i + O(r^{0})$, where $\hat{N}^i$ is the outward unit
normal to the 2-sphere boundary $\partial\Sigma$. Thus, the
relationship between the ADM mass and Komar energy under Weinberg's
asymptoticslly-flat spacetime conditions may be written as
\begin{equation}
M^{\text{ADM}} = E(\xi) - \frac{1}{8 \pi} \int_{\partial\Sigma} 
\bigl( K_{ij} - K \gamma_{ij} \bigr)_{,0}\,
r \hat N^i \hat N^j  dA .
    \label{Result2}
\end{equation}
}

To further simplify the Ricci terms in the integral, we may first
recall two 3+1 decomposition equations of the Einstein field
equations\cite{gourgoulhon2012}
\begin{equation}
R = {}^{(3)}R + K^2 + K_{ij} K^{ij} - \frac{2}{{\cal N}}
\mathcal{L}_{{\cal N} \hat T} K - \frac{2}{{\cal N}} D_i D^i {\cal N} ,
\end{equation}
and 
\begin{equation}
R_{\mu\nu} {\gamma^\mu}_\alpha {\gamma^\nu}_\beta
= -\frac{1}{{\cal N}} \mathcal{L}_{{\cal N}\hat T} K_{\alpha\beta}
- \frac{1}{{\cal N}} D_\alpha D_\beta {\cal N}
+ {}^{(3)}R_{\alpha\beta} + K K_{\alpha\beta}
- 2 K_{\alpha\mu} K^\mu_\beta .
\end{equation}
Thus, the Ricci terms in Eq.~(\ref{Result1}) may be calculated as
\begin{eqnarray}
&&-\frac{1}{8\pi} \int_{\partial\Sigma} (R_{\mu\nu}
- \frac{1}{2} R g_{\mu\nu} ) \hat N^\mu x^\nu dA \nonumber \\
&=& -\frac{1}{8\pi} \int_{\partial\Sigma} ( R_{\mu\nu}
{\gamma^\mu}_\alpha {\gamma^\nu}_\beta \hat{N}^\alpha  x^\beta
- \frac{1}{2} R g_{\mu\nu} \hat N^\mu x^\nu  ) dA \nonumber \\
&=&   -\frac{1}{8\pi} \int_{\partial\Sigma}
\biggl( \Bigl( -\frac{1}{{\cal N}} \mathcal{L}_{{\cal N}\hat T}
K_{\alpha\beta} - \frac{1}{{\cal N}} D_\alpha D_\beta {\cal N}
+ {}^{(3)}R_{\alpha\beta} + K K_{\alpha\beta}
- 2 K_{\alpha\mu} K^\mu_\beta \Bigr) \hat{N}^\alpha  x^\beta \nonumber \\
&&- \frac{1}{2} \Bigl( {}^{(3)}R + K^2 + K_{ij} K^{ij}
- \frac{2}{{\cal N}} \mathcal{L}_{{\cal N} \hat T} K
- \frac{2}{{\cal N}} D_i D^i {\cal N} \Bigr)
g_{\alpha \beta} \hat N^\alpha x^\beta  \biggr) dA
\label{Weinberg1} 
\end{eqnarray}
where from the first to the second line we have used that only the
spatial parts of $R_{\mu\nu}$ contribute to the integral at
spatial infinity.
Recall that $K_{\mu\nu}=O(r^{-2})$ and ${\cal N}\hat T=(1,-\beta^i)$,
Eq.~(\ref{Weinberg1}) may be further simplied as
\begin{eqnarray}
&& -\frac{1}{8\pi} \int_{\partial\Sigma}
\biggl( \Bigl( - K_{\alpha\beta,0}
- D_\alpha D_\beta \text{ln} {\cal N}
+ {}^{(3)}R_{\alpha\beta}  \Bigr) \hat{N}^\alpha  x^\beta
- \frac{1}{2} \Bigl( {}^{(3)}R - 2 K_{,0}
- 2 D_l D^l \text{ln} {\cal N} \Bigr) g_{\alpha \beta}
\hat N^\alpha x^\beta  \biggr) dA  \nonumber \\
&=& -\frac{1}{8\pi} \int_{\partial\Sigma}
\biggl( \Bigl( - K_{ij,0} - D_i D_j \text{ln} {\cal N}
+ {}^{(3)}R_{ij}  \Bigr) \hat{N}^i  x^j
- \frac{1}{2} \Bigl( {}^{(3)}R - 2 K_{,0}
- 2 D_l D^l \text{ln} {\cal N} \Bigr)
\gamma_{ij} \hat N^i x^j  \biggr) dA  \nonumber \\
&=& \frac{1}{8\pi} \int_{\partial\Sigma}
\biggl( \Bigl( K_{ij,0} - K_{,0} \gamma_{ij} \Bigr) 
+ \Bigl( \frac{1}{2}{}^{(3)}R \gamma_{ij} - {}^{(3)}R_{ij}  \Bigr)
+ \Bigl( D_i D_j \text{ln} {\cal N}
-  D_l D^l \text{ln} {\cal N} \gamma_{ij} \Bigr) 
 \biggr) \hat N^i x^j dA \nonumber \\
&=& \frac{1}{8\pi} \int_{\partial\Sigma}
\biggl( \Bigl( K_{ij,0} - K_{,0} \gamma_{ij} \Bigr) 
- {}^{(3)}G_{ij} 
+ \Bigl( D_i D_j \text{ln} {\cal N}
-  D_l D^l \text{ln} {\cal N} \gamma_{ij} \Bigr)  \biggr)
\hat N^i x^j dA
\label{Weinberg2}
\end{eqnarray}
where from the first to the second line, we have choosen the adapted
coordinates system, and ${}^{(3)}G_{ij}$ is the 3-dimensional Einstein
tensor within the hypersurface. Note that the 3-dimensional
Einstein tensor is usually thought to related to the local energy
density and matter stress tensor measured by the Euclerian
observer.\cite{gourgoulhon2012} Since the acceleration of the
Eulerian observer may be defined as\cite{gourgoulhon2012}
$a^i=D^i \text{ln} {\cal N}$, Eq.~(\ref{Weinberg2}) may be further
simplified as
\begin{eqnarray}
&& \frac{1}{8\pi} \int_{\partial\Sigma}
\biggl( \Bigl( K_{ij,0} - K_{,0} \gamma_{ij} \Bigr)
- {}^{(3)}G_{ij} 
+ \Bigl( D_i a_j -  D_l a^l \gamma_{ij} \Bigr) 
 \biggr) \hat N^i x^j dA  \nonumber \\
&=&  \frac{1}{8\pi} \int_{\partial\Sigma}
\biggl( \Bigl( K_{ij,0} - K_{,0} \gamma_{ij} \Bigr) \hat{N}^i  x^j 
- {}^{(3)}G_{ij} \hat{N}^i  x^j - \sigma^{ij} D_i a_j \, r \biggr) dA
\label{Weinberg3}
\end{eqnarray}
where $x^i = r \hat{N}^i + O(r^{0})$ at large $r$ is used, with
$\hat{N}^i$ being the outward unit
normal to the 2-sphere boundary $\partial\Sigma$, and
$\sigma^{ij}\equiv \gamma^{ij}-{\hat N}^i {\hat N}^j$
is the reduced metric on the boundary at spatial infinity.

For the final term in Eq.~(\ref{Weinberg3}), we may calculate as
\begin{eqnarray}
    \frac{1}{8\pi} \int_{\partial\Sigma} \sigma^{ij} D_i a_j  r\, dA
&=& \frac{r}{8\pi} \int_{\partial\Sigma} \sigma^{ij} D_i a_j dA \nonumber \\
&=& \frac{r}{8\pi} \int_{\partial\Sigma} \Bigl( \sigma^{ij} D_i
[(\sigma_{lj} + \hat N_i \hat N_l)\, a^l]
- \sigma^{ij} a^l D_i \gamma_{lj} \Bigr) dA \nonumber \\
&=& \frac{r}{8\pi} \int_{\partial\Sigma}
\Bigl( \sigma^{ij} D_i (\sigma_{lj} a^l) + O(r^{-4}) \Bigr) dA \nonumber \\
&=& \frac{r}{8\pi} \int_{\partial\Sigma}
\Bigl( \mathfrak{D}_A a^A + O(r^{-4}) \Bigr) dA
\end{eqnarray}
where in the first line we assume the boundary $\partial\Sigma$ is
at a large constant radius $r$ with a perturbation in the order
of $o(r)$. In moving from the second line to the third, we assume
$a^i =D^i \ln {\cal N}=O(r^{-2})$ because we are diferentiating a
term from the metric, similarly, we assume $D_j \gamma_{lj}=O(r^{-2})$
for the same reason; and similarly in the next line that
$D_i (\hat N_i \hat N_l) =O(r^{-2})$.
For the final step, we adopt coordinates adapted to the
boundary surface and $\mathfrak{D}_A a^A$ represents the divergence
on the boundary with the indices $A=\{2,3\}$. According to Stokes'
theorem, the final integral vanishes because the boundary of a boundary
is empty. Therefore, Eq.~(\ref{Weinberg3}) reduces into
\begin{eqnarray}
-\frac{1}{8\pi} \int_{\partial\Sigma} (R_{\mu\nu}
- \frac{1}{2} R g_{\mu\nu} ) \hat N^\mu x^\nu dA = \frac{1}{8\pi} \int_{\partial\Sigma}
\biggl( \Bigl( K_{ij,0} - K_{,0} \gamma_{ij} \Bigr) \hat{N}^i  x^j 
- {}^{(3)}G_{ij} \hat{N}^i  x^j \biggr) dA
\label{Weinberg4}
\end{eqnarray}

Inserting Eqs.~(\ref{flat2}) and (\ref{Weinberg4}) into
Eq.~(\ref{Result1}) yields
\begin{equation}
M^{\text{ADM}} = E(\xi) - \frac{1}{8 \pi} \int_{\partial\Sigma} \!\!
{}^{(3)}G_{ij} \hat{N}^i x^j  dA .
    \label{Result2}
\end{equation}
which is the
relationship between the ADM mass and the generalized Komar energy under Weinberg's
asymptotically-flat spacetime conditions.

This completes the proof.\qed

\vskip 0.1in

As an aside, we note that because the Bianchi identity also applies to
the Riemann curvature on the hypersurface, ${}^{(3)}R_{ijkl}$, it
trivially follows by contraction that $D_i {}^{(3)}G^{ij}=0$. This
naively appears to be a statement of momentum conservtion on the
hypersurface.

Equality of the generalized Komar energy and the ADM mass is ensured if
the integral in Eq.~\eqref{Result2} vanishes. A sufficient condition for
this is that the three-dimensional Einstein tensor on the hypersurface,
${}^{(3)}G_{ij}$, decays faster than $r^{-3}$, i.e., ${}^{(3)}G_{ij} =
o(r^{-3})$. Since ${}^{(3)}G_{ij}$ is constructed solely from the
intrinsic metric of the hypersurface and its derivatives, this
constraint applies only to the intrinsic geometry of the spatial slice.
This is a significantly weaker requirement than conditions imposed the
full 4-dimensional Einstein tensor $G_{\mu\nu}=o(r^{-3})$, or
equivalently via the Einstein equations on the stress-energy tensor,
$T_{\mu\nu}=o(r^{-3})$ required in previous work,\cite{Chrusciel1986} or
even $G_{ij}=o(r^{-3})$ required in our York-lite theorem above. Note
that a constraint on $G_{\mu\nu}$, involves both the intrinsic and
extrinsic curvature of the hypersurface. Consequently, by restricting
only the intrinsic geometry, our result for equality relies on a weaker
constraint.


\vskip 0.1in

\end{widetext}

\section{Discussion}

The relationship between the ADM mass and Komar energy in dynamical
spacetimes presents a foundational, yet unresolved, challenge in general
relativity. While the ADM mass offers a well-defined Hamiltonian
approach to total energy for asymptotically-flat dynamical spacetimes
at spatial infinity, and the Komar energy is usually used as a
Noether-charge-based energy for stationary spacetimes, establishing
their relationship in dynamical settings has remained a challenging
issue. In fact, it has often been assumed that no direct relationship
should exist between them in dynamical spacetimes.

This paper confronts this challenge by conducting a rigorous analysis of
the conditions under which the ADM mass and a generalized Komar energy are
equal in dynamical scenarios satisfying a pair of disparate assumptions
about the behavior of asymptotically-flat spacetimes, namely, what
we call the York-lite and Weinberg conditions, given by
Eqs.~(\ref{Yorklite}) and~(\ref{aymp1}), respectively.

\def\lla{
{\color{red} [IS THIS (BELOW) STILL TRUE?]\newline
For a York-lite asymptotically-flat dynamical spacetime, we
demonstrate that the ADM mass equals the generalized Komar energy when the spatial
components of the energy-momentum tensor satisfy the asymptotic
condition $T_{ij}=o(r^{-3})$ and the Killing vector is replaced by an
asymptotically Killing vector that adheres to
${\xi^\mu}_{,00}=o(r^{-3})$ at spatial infinity. In the case of
Weinberg's asymptotically flat dynamical spacetime, the equality between
the ADM mass and Komar energy holds when $T_{ij}=o(r^{-3})$ and
$T_{00}=o(r^{-3})$, with the replacement vector given by
${\xi^\mu}={\cal N} \hat T^\mu + o(r^{-1})$.
}
}

We now turn to the condition for the conservation of the generalized
Komar energy, $E(\xi)$, at spatial infinity. It can be readily shown
that the flux of $J^{\mu}$ through $\Sigma_\infty$ vanishes under
Weinberg's asymptotic conditions. However, for York-lite asymptotic
conditions, this flux can be expressed as
\begin{eqnarray}
\!\!\!    \int_{\Sigma_\infty} \!\!\! J^{\mu} \hat L_\mu
\sqrt{\gamma^{(\partial\Sigma_\infty)}} d^2x
&=& \int_{\Sigma_\infty} \!\!\! {\xi^{[\nu;\mu]}}_{;\nu} \hat L_\mu 
\,dA\,dt \nonumber \\  
    &=& \frac{1}{2} \int_{\Sigma_\infty} \!\!\!
g_{00,0i} g^{00} \hat L_i\, dA\, dt ,
 \label{flux}
\end{eqnarray}
where we have utilized the relation $\hat L^\mu = \hat N^\mu + O(1/r)$.
To ensure the conservation of the generalized Komar energy between
different hypersurfaces, it is necessary that $g_{00,0i}=o(r^{-2})$.
This condition is marginally stronger than the York-like constraint on
the single metric component $g_{00,i}=O(r^{-2})$. However, it is
considerably weaker than Komar's condition for the existence of
an asymptotic Killing vector field, which he argued\cite{komar1962}
required $g_{\mu\nu,0}=o(r^{-2})$. In this regard, our results apply
to spacetimes which though asymptotically-flat fail to be 
asymptotically-stationary at spatial infinity. Our work, therefore,
extends the well-established ADM-Komar equality from
stationary, symmetric spacetimes to a broader, asymptotically-flat 
dynamical context.




\begin{thebibliography}{99}


\bibitem{Wald1984}
\bibinfo{author}{Wald, R. M.}
\newblock \bibinfo{title}{General relativity}.
\bibinfo{pages}{pp.286-295} 
(\bibinfo{year}{University of Chicago press, Chicago and London, 1984}).

\bibitem{Arnowitt1959}
Arnowitt, R., Deser, S.\ \& Misner, C.,
Dynamical Structure and Definition of Energy in General Relativity.
{\it Physical Review} {\bf 116}, 1322-1330 (1959).

\bibitem{Arnowitt1960}
Arnowitt, R., Deser, S.\ \& Misner, C.,
Canonical Variables for General Relativity.
{\it Physical Review} {\bf 117}, 1595-1602 (1960).

\bibitem{Arnowitt1962}
Arnowitt, R., Deser, S.\ \& Misner, C.,
The Dynamics of General Relativity.
In {\it Gravitation: An Introduction to Current Research},
edited by L. Witten, pp. 227-264
(Wiley, New York, 1962).

\bibitem{Bondi62}
Bondi, H., van der Burg, M.  G.  J.  \& Metzner, A. W. K.,
Gravitational Waves in General Relativity. VII.
Waves from Axi-Symmetric Isolated Systems.
{\it Proceedings of the Royal Society of London. Series A,
Mathematical and Physical Sciences} {\bf 269},1336 (1962).

\bibitem{Komar1959}
\bibinfo{author}{Komar, A.},
\newblock \bibinfo{title}{Covariant conservation laws in general relativity}.
\newblock \emph{\bibinfo{journal}{Physical Review}}
  \textbf{\bibinfo{volume}{113}}, \bibinfo{pages}{934-936}
  (\bibinfo{year}{1959}).

\bibitem{beig1978}
\bibinfo{author}{Beig, Robert},
\newblock \bibinfo{title}{Arnowitt-deser-misner energy and g00}.
\newblock \emph{\bibinfo{journal}{Physics Letters A}}
  \textbf{\bibinfo{volume}{69}}, \bibinfo{pages}{153--155}
  (\bibinfo{year}{1978}).

\bibitem{komar1962}
Komar, Arthur,
Asymptotic covariant conservation laws for gravitational radiation.
{\it Physical Review} {\bf 127}, 1411 (1962).

\bibitem{komar1963}
Komar, Arthur,
Positive-definite energy density and global consequences for
general relativity.
{\it Physical Review} {\bf 129}, 1873 (1963).

\bibitem{harte2008}
Harte, Abraham I,
Approximate spacetime symmetries and conservation laws.
{\it Classical and quantum gravity} {\bf 25(20)}, 205008 (2008).

\bibitem{feng2018}
Feng, Justin C,
Some globally conserved currents from generalized Killing vectors
and scalar test fields.
{\it Physical Review D} {\bf 98(10)}, 104035 (2018).

\bibitem{wang2020}
Wang, Zhi-Wei and Braunstein, Samuel L,
Could dark matter be a natural consequence of a dynamical universe?
{\it arXiv preprint}  arXiv:2011.09923 (2020).

\bibitem{wang2021}
Wang, Zhi-Wei and Braunstein, Samuel L,
Noether charge astronomy
{\it arXiv preprint}  arXiv:2105.14985 (2021).



\bibitem{carroll2004}
\bibinfo{author}{Carroll, S. M.}
\newblock \bibinfo{title}{An Introduction
to General Relativity Spacetime and Geometry}.
\bibinfo{pages}{p.251,455-456}, (\bibinfo{year}{Addison Wesley,
San Francisco, 2004}).  


\bibitem{york1979}
\bibinfo{author}{York, J. W.}
\newblock \bibinfo{title}{Kinematics and dynamics of general relativity}.
\newblock \emph{\bibinfo{journal}{Sources of gravitational radiation}}
   \bibinfo{pages}{83--126}
  (\bibinfo{year}{1979}).

\bibitem{gourgoulhon2012}
\bibinfo{author}{Gourgoulhon, E.}
\newblock \bibinfo{title}{3+ 1 Formalism in General Relativity: Bases
of Numerical Relativity}. \bibinfo{pages}{p62, p68, p70, pp79-84},
(\bibinfo{year}{Springer, London, 2012}).

\bibitem{York1980}
\bibinfo{author}{York, J. W.},
\newblock \bibinfo{title}{Energy and momentum of the gravitational field}.
In \emph{\bibinfo{booktitle}{Essays in General Relativity}}
  (\bibinfo{year}{Academic Press, New York, 1980})
\bibinfo{pages}{pp.39-58}.


\bibitem{Weinberg1972}
\bibinfo{author}{Weinberg, S.}
\newblock \bibinfo{title}{Gravitation and cosmology:
principles and applications of the general theory of relativity}.
\bibinfo{pages}{p.167}, (\bibinfo{year}{Wiley, New York, 1972}). 

\bibitem{Chrusciel1986}
\bibinfo{author}{Chru\'{s}ciel, P. T.}
\newblock \bibinfo{title}{A remark on the positive-energy theorem}.
\newblock \emph{\bibinfo{journal}{Classical and Quantum Gravity}}
  \textbf{\bibinfo{volume}{3}}, \bibinfo{pages}{L115}
  (\bibinfo{year}{1986}).


\bibitem{Chrusciel2010}
\bibinfo{author}{Chru\'{s}ciel, P. T.}
\newblock \bibinfo{title}{Lectures on energy in general relativity}.
\newblock  (\bibinfo{year}{2010}).

\bibitem{poisson2004}
\bibinfo{author}{Poisson, E.}
\newblock \bibinfo{title}{A relativist's toolkit: the mathematics
of black-hole mechanics}.
  (\bibinfo{year}{Cambridge university press, 2004}),
\bibinfo{pages}{p13, p15, pp211-212}.





\def \extra{

\bibitem{bertschinger2002}
\bibinfo{author}{Bertschinger, E.}
\newblock \bibinfo{title}{Symmetry transformations, the
Einstein-Hilbert action and gauge invariance}.
\newblock \emph{\bibinfo{journal}{Massachusetts Institute of
Technology, Department of Physics}}
  (\bibinfo{year}{2002}).

\bibitem{palatini1919}
\bibinfo{author}{Palatini, A.}
\newblock \bibinfo{title}{Deduzione invariantiva delle equazioni
gravitazionali dal principio di Hamilton}.
\newblock \emph{\bibinfo{journal}Rendiconti del Circolo Matematico
di Palermo (1884-1940)}
  \textbf{\bibinfo{volume}{43}}, \bibinfo{pages}{203-212}
  (\bibinfo{year}{1919}).
  
\bibitem{Hawking1973}
\bibinfo{author}{Hawking, S. W.} \& \bibinfo{author}{Ellis, G. F. R.}
\newblock \bibinfo{title}{The large scale structure of space-time}.
\bibinfo{pages}{82-83, p.334,}
  (\bibinfo{year}{Cambridge university press, Cambridge, 1973}).  

\bibitem{Parikh2000}
\bibinfo{author}{Parikh, M.\ K.}, \& \bibinfo{author}{Wilczek, F.}
\newblock \bibinfo{title}{Hawking radiation as tunneling}.
\newblock \emph{\bibinfo{journal}{Physical Review Letters}}
  \textbf{\bibinfo{volume}{85}}, \bibinfo{pages}{5042}
  (\bibinfo{year}{2000}).

\bibitem{Cook2000}
\bibinfo{author}{Cook, G.\ B.} 
\newblock \bibinfo{title}{Initial data for numerical relativity}
\emph{\bibinfo{journal}{Living Reviews in Relativity}},
\textbf{\bibinfo{volume}{3}}, \bibinfo{pages}{5} (\bibinfo{year}{2000}).

\bibitem{Brill1963}
\bibinfo{author}{Brill,\ D.\ R.}, \& \bibinfo{author}{Lindquist,\ R.\ W.}
\newblock \bibinfo{title}{Interaction energy in geometrostatics}.
\newblock \emph{\bibinfo{journal}{Physical Review}}
  \textbf{\bibinfo{volume}{131}}, \bibinfo{pages}{471}
  (\bibinfo{year}{1963}).


\bibitem{Badri2002}
\bibinfo{author}{Krishnan, B.}
\newblock \bibinfo{title}{Isolated Horizons in Numerical Relativity},
\newblock \emph{\bibinfo{journal}{Ph.D. Thesis}}
(The Pennsylvania State University, 2002), pp.~68.

\bibitem{jacobson1995}
\bibinfo{author}{Jacobson, T.}
\newblock \bibinfo{title}{Thermodynamics of spacetime: The Einstein
equation of state}.
\newblock \emph{\bibinfo{journal}{Physical Review Letters}}
  \textbf{\bibinfo{volume}{75}}, \bibinfo{pages}{1260}
  (\bibinfo{year}{1995}).

\bibitem{Hawking1975}
\bibinfo{author}{Hawking, S.\ W.}
\newblock \bibinfo{title}{Particle creation by black holes},
\emph{\bibinfo{journal}{Communications in Mathematical Physics}}
\textbf{\bibinfo{volume}{ 43}}, \bibinfo{pages}{199-220}
 (\bibinfo{year}{1975}).

\bibitem{Bardeen1973}
\bibinfo{author}{Bardeen, James M and Carter, Brandon and Hawking, Stephen W}
\newblock \bibinfo{title}{The four laws of black hole mechanics},
\emph{\bibinfo{journal}{Communications in mathematical physics}}
\textbf{\bibinfo{volume}{31}}, \bibinfo{pages}{161--170}
 (\bibinfo{year}{1973}).


\bibitem{hawking1968}
Hawking, Stephen W,
Gravitational radiation in an expanding universe.
{\it Journal of Mathematical Physics} {\bf 9}, 598--604 (1968).

\bibitem{Penrose82}
Penrose, R.,
Quasi-local mass and angular momentum in general relativity.
{\it Proceedings of the Royal Society of London. Series A,
Mathematical and Physical Sciences} {\bf 381}, 53-63 (1982).

\bibitem{wald2000}
Wald, Robert M and Zoupas, Andreas,
General definition of “conserved quantities” in general relativity
and other theories of gravity.
{\it Physical Review D} {\bf 61}, 084027 (2000).

\bibitem{harlow2020}
Harlow, Daniel and Wu, Jie-qiang,
Covariant phase space with boundaries.
{\it Journal of High Energy Physics} {\bf 2020}, 1--52 (2020).

\bibitem{Szabados09}
Szabados, L.\ B.,
Quasi-Local Energy-Momentum and Angular Momentum in General Relativity.
{\it Living Reviews in Relativity} {\bf 12}, 1-163 (2009).

}
  
\end{thebibliography}
\end{document}